\documentclass[twocolumn,revtex4,apj,iop, twocolappendix, numberedappendix]{openjournal}
\usepackage{xcolor}
\usepackage{graphicx}
\usepackage{amsfonts}
\usepackage{amssymb}
\usepackage{url}
\usepackage[breaklinks,colorlinks,citecolor=blue,linkcolor=blue,urlcolor=blue]{hyperref}
\usepackage{amsmath}
\usepackage[varg]{txfonts}
\usepackage[normalem]{ulem}
\usepackage{fontawesome}
\DeclareMathAlphabet{\mathbbold}{U}{bbold}{m}{n}

\renewcommand{\d}{\mathrm{d}}
\newcommand{\p}{\partial}

\newcommand{\e}{\mathrm{e}}

\newcommand{\s}{\mathrm{s}}
\renewcommand{\o}{\mathrm{o}}

\setcitestyle{numbers,square}
\begin{document}

\title{Redshift Drift in Relativistic N-body Simulations}

\author{Alexander Oestreicher$^{\;a,}$\footnote{alexo@cp3.sdu.dk}}
\author{Chris Clarkson$^{\;b,\,c}$}
\author{Julian Adamek$^{\;d}$}
\author{Sofie Marie Koksbang$^{\;a}$ \vspace{0.15cm}}

\affiliation{${}^a$ CP3-Origins, University of Southern Denmark, Campusvej 55, DK-5230 Odense M, Denmark}
\affiliation{${}^b$ Department of Physics \& Astronomy, Queen Mary University of London,  Mile End Road, London E1 4NS, UK\vspace{0.05cm}}
\affiliation{${}^c$ Department of Physics \& Astronomy, University of the Western Cape, Cape Town 7535, South Africa\vspace{0.05cm}}
\affiliation{${}^d$ Institut f\"ur Astrophysik, Universit\"at Zürich, 	Winterthurerstrasse 190, 8057 Z\"urich, Switzerland \vspace{0.15cm}}

\begin{abstract}
	The cosmological redshift drift promises to be the first observable directly measuring the evolution of the cosmic expansion rate and should be detectable with upcoming surveys by the Square Kilometre Array and the Extremely Large Telescope. To prepare for these upcoming measurements we study the redshift drift in detail using the relativistic N-body code \texttt{gevolution}, focusing on inhomogeneity-induced fluctuations. Using a ray-tracer, we calculate the redshift drift directly from the light cone at two different time steps. To investigate observer-dependent biases we consider 10 different observers. We find that inhomogeneity-induced fluctuations in the redshift drift can in extreme cases be of the same order as the cosmic signal for $z\lesssim0.15$. By comparing our results to first-order perturbation theory, we find that the extreme outliers are due to peculiar motion in over-densities and can be described by first-order perturbation theory to percent precision. We calculate angular power spectra that fit very well with our predictions based on perturbation theory at linear scales and show a surprisingly large non-linear signal. This shows that redshift drift not only has the power to measure the background expansion, but could also deliver information about the velocity and acceleration fields in clusters.
\end{abstract}

\maketitle 
\tableofcontents

\section{Introduction}
One of the primary tasks in cosmology is to measure the cosmic expansion rate. Such measurements constrain the spacetime geometry, the energy content of our universe, and the theory of gravity describing its evolution. Measurements of the expansion rate through supernovae of type Ia gave us the first hint that our universe is expanding in an accelerated manner, thereby prompting us to include dark energy in our model of the Universe \cite{Riess1998, Perlmutter1999}. However, so far, most of our measurements of the expansion rate have been indirect. Measurements of supernovae for example constrain the distance-redshift relation and therewith an integrated measure of the expansion rate. A direct measurement of the cosmic expansion rate and its evolution would be a huge step forward in constraining and understanding our universe's evolution and the phenomenology of dark energy.

Such a direct measurement of the cosmic expansion and its evolution is for the first time becoming feasible with the current generation of telescopes. Telescopes such as the Square Kilometre Array (SKA) and the Extremely Large Telescope (ELT) are predicted to be able to measure the real-time variation of the cosmological redshift \cite{Liske2008, Kloeckner2015, Cooke2020, Rocha2023, Trost2025}, the so-called redshift drift \cite{Sandage1962, McVittie1962}. The redshift drift is the variation of the redshift, $z$, of a given source observed at two different times $\delta t_\o$ apart, i.e.
\begin{equation}
	\delta z :=\frac{dz}{dt_\o} \delta t_\o= z(t+\delta t)-z(t)\;,
	\label{eq:dz_def}
\end{equation}
where $\delta t$ is the proper time interval at the source corresponding to the time interval $\delta t_\o$ at the observer.
In an FLRW universe the redshift drift is related to the cosmic expansion rate $E=H(z)/H_0$ as 
\begin{equation}
	\delta z = \delta t_\o H_0(1+z-E(z))\;,
\end{equation}
where $H_0$ is the Hubble constant at the time of observation. Therefore, up to a constant prefactor, $E(z)$ can be directly constrained. 

The real universe is, however, only homogeneous and isotropic, i.e. described by an FLRW metric, if viewed on the largest scales. On smaller scales, the Universe is full of structures. These structures affect any cosmological observables, including the redshift drift. Indeed, the effects of structures have been identified as one of the main obstacles for obtaining unbiased redshift drift measurements \cite{Liske2008, Cooke2020}. It was estimated in \cite{Liske2008, Cooke2020}, that the noise will be subdominant to the signal expected for observations with the ELT unless based on observations of a small number of objects with large peculiar velocity and acceleration. Furthermore, observations with the SKA will average over billions of galaxies \cite{Kloeckner2015}, meaning that peculiar motion should only introduce random noise and not bias the result. It remains important to precisely quantify the size of the noise in a realistic simulation. This is both important in order to confirm that the measurements will indeed be unbiased and do not suffer from observer-dependent effects. Furthermore, in addition to being a source of noise in the background signal, the peculiar motion of objects might provide a useful signal in itself. For example \cite{Amendola2008} suggested that the peculiar acceleration of objects in nearby clusters could be used to map out their gravitational potential. It is therefore interesting to study whether such signals can be extracted from redshift drift measurements.
The time variation of the redshift, in particular due to peculiar motion, could also be important for future gravitational wave detectors such as LISA or DECIGO \cite{Bonvin2017, Seto2001}.
\\ \\ 
The fluctuations in the redshift drift have already been studied in several cosmological simulations. In \cite{Koksbang2023}, the redshift drift was studied using the Newtonian N-body code GADGET-2 \cite{Springel2005_GADGET2}, while \cite{Koksbang2024} studied the redshift drift in fully relativistic simulations of structure formation run with the Einstein Toolkit \cite{Macpherson2017,Macpherson2019}. Both papers considered an Einstein-de Sitter universe and calculated the redshift drift using the general expansion introduced in \cite{Heinesen2021} keeping leading-order terms. Furthermore, both studies used quite low resolution\footnote{The resolution of the actual simulation used in \cite{Koksbang2023} is not particularly low with $512^3$ particles on a 512 Mpc/h grid. However, redshift drift computations were produced only after smoothing the data onto the grid.} which may be important when assessing the size of the effects peculiar motion has on redshift drift fluctuations. In \cite{Bessa2024}, the redshift drift fluctuations were studied using GADGET-4 \cite{Springel2021_GADGET4}, calculating the redshift drift from perturbation theory on the spatial background hypersurfaces/snapshots of the simulation. This study was the first to consider redshift drift fluctuations in a simulation based on a $\Lambda$CDM background, but the study only estimated the redshift drift signal on constant coordinate-time snapshots and thus at background redshifts rather than using the fluctuating redshift. In addition, the study compared to linear perturbation theory and found a striking difference between perturbative predictions and simulation results. 

The goal of this paper is to address the shortcomings of these earlier studies by computing the exact redshift drift in a $\Lambda$CDM-based relativistic N-body simulation with high resolution, with the aim of understanding the relationship between simulation and perturbative results. To do this, we use the relativistic N-body code \texttt{gevolution} \cite{Adamek2016_Nat}. 
\\ \\ 
The paper is structured as follows. We start in Sect.~\ref{sec:sim_setup} by introducing our simulation setup and explaining how we calculate the redshift drift. Next, we derive the redshift drift and the redshift drift fluctuation power spectrum in a perturbed FLRW universe in Sect.~\ref{sec:theory}. In Sect.~\ref{sec:results} we present the results of our simulation and compare them with the results from Sect.~\ref{sec:theory} and other results from the literature. In Sect.~\ref{sec:obs_comparison} we discuss how our results relate to current and future observational efforts, before concluding in Sect.~\ref{sec:conclusions}.

\section{Simulation Setup}
\label{sec:sim_setup}
The results presented in this paper are produced with the relativistic N-body code \texttt{gevolution}\footnote{\url{https://github.com/gevolution-code}}\cite{Adamek2016_Nat}. \texttt{gevolution} employs a particle-mesh approach and implements the perturbed FLRW metric in the Poisson gauge 
\begin{align}
	\d s^2 = a^2(\tau)\big[&-\e^{2\Psi}\,\d\tau^2-2B_i\,\d x^i\d\tau \\ \nonumber 
	&+\big(\e^{-2\Phi}\,\delta_{ij}+h_{ij}\big)\,\d x^i \d x^j \big]\;,
\end{align}
where $\tau$ is the conformal time, $a$ the background scale factor and $x^i$ are co-moving coordinates. Here, $\Psi$, $\Phi$ are the two scalar perturbations, known as the Bardeen potentials, and $B_i$, $h_{ij}$ are the vector and tensor perturbations. The metric components are stored on a regular Cartesian grid and evolved according to the field equations of general relativity expanded in a weak field expansion. For the exact details of the expansion and an overview of the terms that are kept, the reader is referred to \cite{Adamek2016}. The positions and velocities of the particles are updated by solving the geodesic equation. 

In addition to the metric on the Cartesian grid and full particle positions and velocities, \texttt{gevolution} can output particle positions and metric fluctuations on background light cones for an observer at a given position and observation time. In this case, the metric is stored on a \text{HEALPix} \cite{HEALPix} grid. 

One can post-process the output with a raytracer to compute the redshift and position of the individual particles on the sky. We use the raytracer first introduced in \cite{Adamek2019}. The raytracer uses an iterative procedure, shooting light rays from the observer positions towards the source, solving the geodesic equation along the ray using the perturbed metric, though only retaining the scalar perturbations. If the ray misses the source, the initial shooting angle is adjusted for the next pass until the correct geodesic is calculated.

To study the redshift drift we simulate the light cones for 10 different observers, randomly placed throughout the simulation volume. Each light cone consists of a full sky section with radius $0.4\times \mathrm{boxsize}$ and a pencil beam that extends to $1.05\times\mathrm{boxsize}$ with an opening half-angle of $25^\circ$. The observers look along a diagonal of the simulation box, chosen randomly from the eight possible diagonal directions. This setup allows us to go to the largest possible redshift, without having any part of the simulation appear on the light cone more than once. To create the light cone, the periodic boundary conditions of the simulation are used. 

For each observer, we simulate two light cones at two different observation times $\tau_1,\tau_2$, ensuring that the two observation times are separated by at least one simulation time step. For particles that appear on both light cones, the redshift drift is then simply given as the difference between the particles redshift on the two light cones rescaled to the proper observation time interval $\delta t_\o$ according to\footnote{We use the coordinate time $t$ in the background spacetime here, assuming the difference to the proper time to be very small.}
\begin{equation}
	\delta z = \frac{z(\tau_1)-z(\tau_2)}{t(\tau_1)-t(\tau_2)} \delta t_\o\;.
\end{equation}
 We chose $\delta t_\o = 30$ years since this is comparable to the expectation for surveys with the SKA and ELT and is also commonly used in earlier studies of the redshift drift, making comparison easier. As it is only a constant prefactor, all results can easily be scaled to another choice of $\delta t_\o$ if desired.

 With this setup, we have chosen observers co-moving with the expansion of the universe. In a realistic scenario, the observer will be moving with respect to the background. We neglect such motion here as we are interested in the effects of the large scale structure of the universe. The contribution to the redshift drift signal from our local motion has been studied earlier \cite{Bolejko2019, Inoue2020} and we expect its contributions to the observed redshift drift signal can be removed.

Our simulation volume has a co-moving boxsize of $1024$ Mpc/h, contains $1024^3$ particles, and has $1024^3$ grid cells on which the metric is sampled. To get results without numerical artefacts, a low Courant factor has to be used. We found that $c=\Delta\tau/\Delta x=3$ was sufficient for our purposes. The two light cones were output for $\tau_1=\tau(z=0)$ and $\tau_2=\tau(z=0.002)$, which was the smallest possible separation that had at least one simulation timestep between the two times. 

Our simulation uses a flat $\Lambda$CDM model as background, with initial conditions created using \texttt{CLASS}\footnote{\url{http://class-code.net}}\ \cite{CLASS_II}, using the following cosmological parameters $h = H_0 / (100 \,\mathrm{km}\,\mathrm{s}^{-1}\mathrm{Mpc}^{-1}) = 0.67556,\; \Omega_{c}h^2= 0.12038,\; \Omega_{b}h^2= 0.022032,\; A_s = 2.215\times 10^{-9}$ (at $0.05\,\mathrm{Mpc}^{-1}$) and $n_s= 0.9619$, which are the standard parameters for \texttt{gevolution} and very close to standard Planck parameters \cite{Planck2018_Parameters}.

With these choices, the particles in our simulation have a mass of $m \approx 1.28\times 10^{11}\,M_\odot$, roughly an order of magnitude lighter than the Milky Way.

\section{Analytical Predictions}
\label{sec:theory}
The redshift drift in a perturbed FLRW universe was first discussed in \cite{Linder2010,Kim2015,Marcori2018} and later derived in \cite{Bessa2023} in a gauge-independent manner. In this paper, we will stick closest to the result by \cite{Marcori2018}, but extend it by a term that appears when observing on the light cone. We repeat the main steps of the calculation to make this paper self-contained and introduce our notation. We will only consider the dominant corrections to the redshift and redshift drift which come from the peculiar motion and not the metric fluctuations. In the derivations below we therefore assume that the wave vector is that of an FLRW spacetime and that the only redshift corrections from inhomogeneities are those from the peculiar motion of the source and observer which will be taken into account at the linear level.
We thus assume that the light paths including the null tangent vector are in accordance with an FLRW spacetime with line-element
\begin{equation}
	\d s^2 = - \d t^2 + \delta_{ij}\;a^2\d x^i \d x^j\;,
\end{equation}
where $t$ is cosmic time, $a$ the scale factor and $x^i$ are co-moving coordinates. We have set $c=1$ for convenience.

\subsection{Redshift Drift}
The redshift is defined via the observed relative frequency difference between emission and observation  
\begin{equation}
	z = \frac{\omega^{obs}_\s}{\omega^{obs}_\o}-1\;.
\end{equation}
A moving observer with 4-velocity $u^\mu$ measures 
\begin{equation}
	\omega^{obs} = -u_\mu k^\mu\;.
	\label{eq:obs_w}
\end{equation}
Within the geometric optics approximation the wave vector has to be $k^\mu = \omega\; (1,\hat k^i)$ \cite{Gravitation}, where $\omega$ is the angular frequency measured by a local Lorentz observer and $\hat k$ is the unit direction vector pointing in the direction of propagation of the wave. Considering an observer with 4-velocity $u^\mu = (1,u^i)$ one finds
\begin{equation}
	\omega^{obs} = \omega\; (1-u_i \hat k^i)\;,
\end{equation}
where $u^i=\d x^i/\d t$ is the peculiar velocity with respect to the Hubble flow. With this, the redshift for a moving source and observer in a FLRW universe becomes
\begin{align}
	1 + z &= \frac{\omega_\s (1-u_i \hat k^i\vert_\s)}{\omega_\o (1-u_i \hat k^i\vert_\o)}\;.
\end{align}
Assuming the peculiar velocity to be small and expanding to first order we find
\begin{align}
	1+z\approx \frac{\omega_\s}{\omega_\o}(1-u_i \hat k^i\vert_\s+u_i \hat k^i\vert_\o)\;.
\end{align}
From the geodesic equation we know that $a\omega$ is constant along light rays in FLRW spacetimes, allowing us to replace $\omega$ and find 
\begin{equation}
	1+z\approx \frac{a_\o}{a_\s}(1-u_i \hat k^i\vert_\s+u_i \hat k^i\vert_\o)\;.
\end{equation}
We now replace $u^i$ and $\hat k^i$ with their co-moving equivalents $v^i=\d x^i/\d \tau=a u^i$ and $n^i=a \hat k^i$, evaluate the scalar product, and introduce the notation $\bold{a}\cdot\bold{b} = \delta_{ij}a^i b^j$ for the Euclidian scalar product. This leaves us with the simple expression
\begin{equation}
	1+z\approx \frac{a_\o}{a_\s}\left[1+\left(\bold v_\o-\bold v_\s\right) \cdot\bold n\right]\;,
	\label{eq:redshift}
\end{equation}
where we also used that $\bold n$ is constant along null geodesics in an FLRW universe. An observer a small amount of time $\delta t$ later will instead observe
\begin{equation}
	1+z+\delta z \approx \frac{a_\o(t+\delta t)}{a_\s (t+\delta t)}\left[1+\left(\bold v_\o(t+\delta t)-\bold v_\s(t+\delta t)\right)\cdot \bold n\right]\;,
\end{equation}
where we introduced the redshift drift $\delta z = z(t+\delta t)-z(t)$ and again used that $\bold n $ remains constant along the null geodesic. We can rewrite this as 
\begin{equation}
	1+z+\delta z \approx \frac{a_\o+\dot a_\o \delta t_\o}{a_\s +\dot a_\s \delta t_\s}\left[1+\left(\bold v_\o+\dot{\bold v}_\o\delta t_\o-\bold v_\s-\dot{\bold v}_\s\delta t_\s\right)\cdot \bold n \right]\;.
\end{equation}
Keeping only terms first order in $\delta t$ and $\bold v$, subtracting \eqref{eq:redshift} and solving for the redshift drift leaves us with
\begin{equation}
	\delta z = \delta t_\o\left[(1+z)\left(H_\o+\dot{\bold v}_\o\cdot \bold n\right)-\left(H_\s+\dot{\bold v}_\s\cdot\bold n \right)\right]\;,
\end{equation}
where we used that $\delta t_\o/\delta t_\s=1+z$ and that $(1+\bar z)\dot{\bold v}=(1+z)\dot{\bold v}$ at first order. This is not yet the observable expression. We still have to account for two things: Firstly, an observer does not see the direction $\bold n$ in which the wave is propagating, but rather a direction $\bold e$ on the sky. We can decompose the co-moving wave vector $n^\mu=a k^\mu$ as follows
\begin{equation}
	n^\mu = \omega_{obs}(v^\mu-e^\mu)\;,
\end{equation}
where the minus sign is chosen such that $e^\mu$ points from the observer to the source and the prefactor follows from \eqref{eq:obs_w} and the fact that $k^\mu$ is null. One can show that 
\begin{equation}
	e^i = -n^i+\perp^i_j v^j
	\label{eq:e_n}
\end{equation}
with the projector 
\begin{equation}
	\perp^i_j = \delta^i_j - n^i n_j\;.
\end{equation}
The vector $\bold n$ only ever appears doted with $\bold v$. We can therefore to first order replace $\bold n = - \bold e$, yielding
\begin{equation}
	\delta z = \delta t_\o\left[(1+z)\left(H_\o-\dot{\bold v}_\o\cdot \bold e\right)-\left(H_\s-\dot{\bold v}_\s\cdot\bold e \right)\right]\;.
    \label{eq:dz_macori}
\end{equation}
Since $\bold n$ is constant along the geodesic, we are free to choose it at any point along the light ray. Since we further replaced $\bold n =-\bold e$ we are now free to choose $\bold e$ at any point. We will always choose $\bold e_\o$ later, but simply keep writing $\bold e$ for ease of notation.
The result \eqref{eq:dz_macori} now agrees with the result in \cite{Marcori2018}, but is not yet the one we observe on the light cone. The second effect we have to account for comes about from the following: So far all quantities are evaluated at the background time $t$, related to hypersurfaces of constant background redshift $\bar z$. Observationally we do not have access to $\bar z$, only to the observed redshift $z=\bar z + \Delta z$. For any function we can relate
\begin{equation}
	f(t(\bar z))\equiv f(\bar z) = f(z-\Delta z)\approx f(z)-\frac{\d f}{\d \bar z}\Delta z\;,
\end{equation}
\cite{Bonvin2011,Bessa2023}. Within perturbation theory, such corrections need only be applied to background quantities, i.e.\ in our case only $H_\s$. We have 
\begin{align}
	H_\s(\bar z)=H_\s(z -\Delta z)&=H_\s(z)-\frac{\d H_\s}{\d \bar z}\Delta z \nonumber \\
	&=H_\s(z)-\frac{\d H_\s}{\d t}\frac{\d t}{\d a_\s}\frac{\d a_\s}{\d \bar z}\Delta z \nonumber \\
	&=H_\s(z)-\frac{\dot H_\s}{\dot a_\s}\frac{\d}{\d \bar z}\left(\frac{a_\o}{1+\bar z}\right)\Delta z \nonumber \\
	&=H_\s(z)+\frac{\dot H_\s}{H_\s}\frac{a_\o}{a_\s}\frac{1}{(1+\bar z)^2}\Delta z \nonumber \\
	&=H_\s(z)+\frac{\dot H_\s}{H_\s}(\bold v_\s-\bold v_\o)\cdot \bold e\;,
\end{align}
where we inserted $\Delta z = a_\o/a_\s (\bold v_\s-\bold v_\o)\cdot \bold e$ as can be read of from \eqref{eq:redshift}, again using $\bold n=-\bold e$. With this, our final result for the redshift drift is 
\begin{equation}
	\delta z = \delta t_\o\left((1+z)\left(H_\o-\dot{\bold v}_\o \cdot\bold e\right)-H_\s-\frac{\dot H_\s}{H_\s}(\bold v_\s-\bold v_\o)\cdot \bold e+\dot{\bold v}_\s \cdot\bold e\right)\;,
\end{equation}
which for an observer co-moving with the cosmic expansion reduces to 
\begin{equation}
	\delta z = \delta t_\o\left((1+z)H_\o-H_\s-\frac{\dot H_\s}{H_\s}\bold v_\s\cdot \bold e+\dot{\bold v}_\s\cdot \bold e\right)\;.
	\label{eq:dz_v1}
\end{equation}
The first two terms appearing here is simply the FLRW background result $\delta\bar z = \delta t_\o((1+z)H_\o-H_\s)$. Additionally we find contributions by the peculiar velocity and acceleration. For later convenience we introduce the shorthands
\begin{equation}
	\delta z^{v} = -\delta t_\o \frac{\dot H_\s}{H_\s}\bold v_\s\cdot \bold e\;, \quad \delta z^{\dot v} = \delta t_\o \dot{\bold v}_\s\cdot \bold e\;.
	\label{eq:peculiar_vel_acc}
\end{equation}
We can also rewrite \eqref{eq:dz_v1} as
\begin{align}
	\delta z = \delta t_\o\left((1+z)H_\o-H_\s+H_\s\frac{\d}{\d t}\left( \frac{\bold v_\s}{H_\s}\right)\cdot \bold e\right)\;, 
	\label{eq:dz_v2}
\end{align}
which will come in handy when deriving the angular power spectrum in the next subsection. Eq. \eqref{eq:dz_v1} agrees with the expression obtained in \cite{Bessa2023}, except for the observer-dependent term that was neglected there, and minus all terms arising from metric fluctuations, which we neglect here. 

\subsection{Angular Power Spectrum}
\label{sec:theory_ps}
We now derive an expression for the angular power spectrum $C_l$, defined via 
\begin{equation}
	\delta_{ll'}\delta_{mm'}C_l=\langle a_{lm}a^*_{l'm'}\rangle
	\label{eq:Cl_def}
\end{equation}
for the redshift drift fluctuations 
\begin{equation}
	\Delta\delta z = \frac{\delta z - \delta \bar z}{\delta \bar z} = \frac{H_\s}{(1+z)H_\o-H_\s}\frac{\d}{\d t}\left( \frac{\bold v_\s}{H_\s}\right)\cdot \bold e\;.
    \label{eq:rsd_fluctuations}
\end{equation}

We note that the power spectrum of redshift drift fluctuations has already been calculated to first order in perturbation theory in \cite{Bessa2023} in a fully gauge-independent manner, accounting not only for the velocity perturbations but also for metric perturbations. Since we choose a different convention for $\Delta\delta z$ (\cite{Bessa2023} neglect the $H_0$ term, which we retain), we present our own derivation here.
\\ \\
If we observe $\Delta\delta z(\bold e, z)$ on the sky, we can expand into spherical harmonics and the $a_{lm}$ are given by 
\begin{equation}
	a_{lm}^{\Delta\delta z}(z) = \int \d\Omega_{\bold e}\; \Delta\delta z(\bold e, z)Y_{lm}^*(\bold e)\;.
\end{equation}
Making use of the 3d-Fourier Transform, we can write
\begin{equation}
	a_{lm}^{\Delta\delta z}(z) = \frac{1}{(2\pi)^3}\int \d\Omega_{\bold e}Y_{lm}^*(\bold e)\int \d^3 k\; \Delta\delta z(\bold k, z)\, \e^{-i r \bold k \cdot \bold e }\;,
\end{equation}
where  $r$ is the co-moving distance to our source. Introducing the velocity potential $v$ such that $ \bold v = -\bold \nabla v$, we simply have $\bold v = iv\,\bold k$ in Fourier space. Using this we can write the redshift drift fluctuation in Fourier space as 
\begin{equation}
	\Delta\delta z (\bold k,z)= \frac{H_\s}{(1+z)H_\o-H_\s}\frac{\d}{\d t}\left( \frac{ v_\s}{H_\s}\right)i\,\bold k \cdot \bold e\;.
\end{equation}
To find the time derivative of the velocity potential we make use of the continuity equation
\begin{equation}
	k^2 v = - a \dot \delta(\bold k, t)\;.
\end{equation}
Using that $\delta(\bold k, t)=D_+(t)\delta_0(\bold k)$ within linear perturbation theory (where $D_+$ is the linear growth factor) we have 
\begin{equation}
	v = -\frac{1}{k^2}a f H D_+ \delta_0(\bold k)\;, 
\end{equation}
where we introduce the growth rate $f=\d\ln D_+/\d\ln a$. With this we now easily find 
\begin{align}
	\frac{\d}{\d t}\left(\frac{v}{H}\right)&=-(HafD_++a^2Hf'D_++aHf^2D_+)\frac{1}{k^2}\delta_0(\bold k) \nonumber \\
	&=-\left(1+f+\frac{\d\ln f}{\d\ln a}\right)\frac{HafD_+}{k^2}\delta_0(\bold k)\;.
\end{align}
Putting everything together we have 
\begin{equation}
	\Delta\delta z (\bold k,z)= -T(z)\,\delta_0(\bold k)\frac{i}{k}\bold{\hat k} \cdot \bold e\;,
\end{equation}
where we defined
\begin{equation}
	T(z) = \frac{H_\s}{(1+z)H_\o-H_\s}\left(1+f+\frac{\d\ln f}{\d\ln a}\right)HafD_+\Big\vert_\s\,.
\end{equation}
Returning to the $a_{lm}$ we now have 
\begin{align}
	a_{lm}^{\Delta\delta z}(z) =& -\frac{1}{(2\pi)^3}\int \d\Omega_{\bold e}Y_{lm}^*(\bold e)\int \d^3 k\; T(z)\,\delta_0(\bold k)\frac{i}{k}\bold{\hat k} \cdot \bold e\; \e^{-i r \bold k \cdot \bold e } \nonumber \\
	=& \frac{1}{(2\pi)^3}\int \d\Omega_{\bold e}Y_{lm}^*(\bold e)\int \d^3 k\; T(z)\,\delta_0(\bold k)\frac{1}{k}\p_{kr} \e^{-i r \bold k \cdot \bold e }\;.
\end{align}
Expanding $\exp(-i r \bold k \cdot \bold e)$ into plane waves using the identity
\begin{equation}
	\e^{-i r \bold k\cdot\bold e} = 4\pi \sum_{l=0}^\infty\sum_{m=-l}^l (-i)^l j_l(kr) Y_{lm}(\bold e)Y_{lm}^*(\hat{\bold k})\;,
\end{equation}
and making use of the orthonormality of the spherical harmonics to evaluate the angular integral leads to
\begin{equation}
	a_{lm}^{\Delta\delta z}(z) = \frac{(-i)^l}{2(\pi)^2}\int \d^3 k\; \frac{1}{k}T(z)\,\delta_0(\bold k) j'_{l}(kr_s)Y_{lm}^*(\hat{\bold k})\;.
\end{equation}
Here $j'_l$ is the first derivative of the spherical Bessel function with respect to its argument. Inserting this result into \eqref{eq:Cl_def}, using the definition of the density fluctuation power spectrum
\begin{equation}
	\langle \delta(\bold k)\delta^*(\bold k')\rangle = (2\pi)^3 \delta (\bold k -\bold k')P_\delta(k)\;,
\end{equation}
integrating over the delta function and using the orthonormality of the spherical harmonics a second time to integrate over the angular part of the $k$ integral one finds
\begin{equation}
	C_l^{\Delta\delta z}(z) = \frac{2}{\pi}\int \d k \; {j'_l}^2(kr)T^2(z)P_\delta^0(k)\;.
\end{equation}
Here $P_\delta^0$ is the density fluctuation power spectrum today. In a real survey we cannot observe one precise redshift but rather consider a bin around the redshift $z\pm \d z$. We can account for this by adding a window function \cite{CLASSgal}
\begin{equation}
	C_l^{\Delta\delta z}(z) = \frac{2}{\pi}\int \d k\; P_\delta^0(k) \left(\int \d z\; W(z) j'_l(kr(z))T(z)\right)^2\;.
	\label{eq:rsd_Cl}
\end{equation}
We will mainly use a normalized tophat
\begin{equation}
	W(z) = 
	\begin{cases} 
		1/(2\d z) & \text{if} \quad z-\d z \leq z \leq z+\d z \\
		0         & \text{otherwise}
	\end{cases}\:.
\end{equation}
As we will see later the choice of the window function and its width has a large influence on the resulting power spectra. 

To evaluate the expression \eqref{eq:rsd_Cl}, we use CLASS to calculate the linear matter power spectrum today $P_\delta^0(k)$ with the same cosmological parameters listed in Sect.~\ref{sec:sim_setup} and use \texttt{FFTlog-and-beyond}\footnote{\url{https://github.com/xfangcosmo/FFTLog-and-beyond}}\cite{Fang_2020_FFTlog} a library specifically designed to handle integrals over spherical Bessel functions and their derivatives. We use \texttt{FFTlog-and-beyond} to implement the first integral over the redshift $z$, transforming it to an integral over the co-moving distance $r$ to do so, and then use the composite Simpson's rule to evaluate the remaining integral over $k$. Our implementation can be found here \href{https://github.com/oestreichera/redshiftdrift_cl}{\faicon{github}}.
\\ \\
As mentioned earlier, the angular power spectrum of redshift drift fluctuations has been presented in \cite{Bessa2023} in a different manner than the current presentation. The authors of \cite{Bessa2023} derived the power spectrum in a fully gauge-independent manner and also included metric fluctuations, which we neglected here. According to the results we will show in Sect.~\ref{sec:results_theor_pred} and also the conclusions in \cite{Bessa2023}, neglecting the metric perturbations is very well justified and should not influence the result significantly. Nevertheless, our results disagree with those of \cite{Bessa2023}, especially for the low redshifts important in this paper. We will see that our result agrees very well with our simulation results and also with the simulation results by \cite{Koksbang2024} and are therefore inclined to trust it. A full discussion of the differences between our results and those of \cite{Bessa2023} can be found in Appx.~\ref{appendix:Bessa2023}.

\begin{figure}
	\centering
	\includegraphics[width=0.85\linewidth]{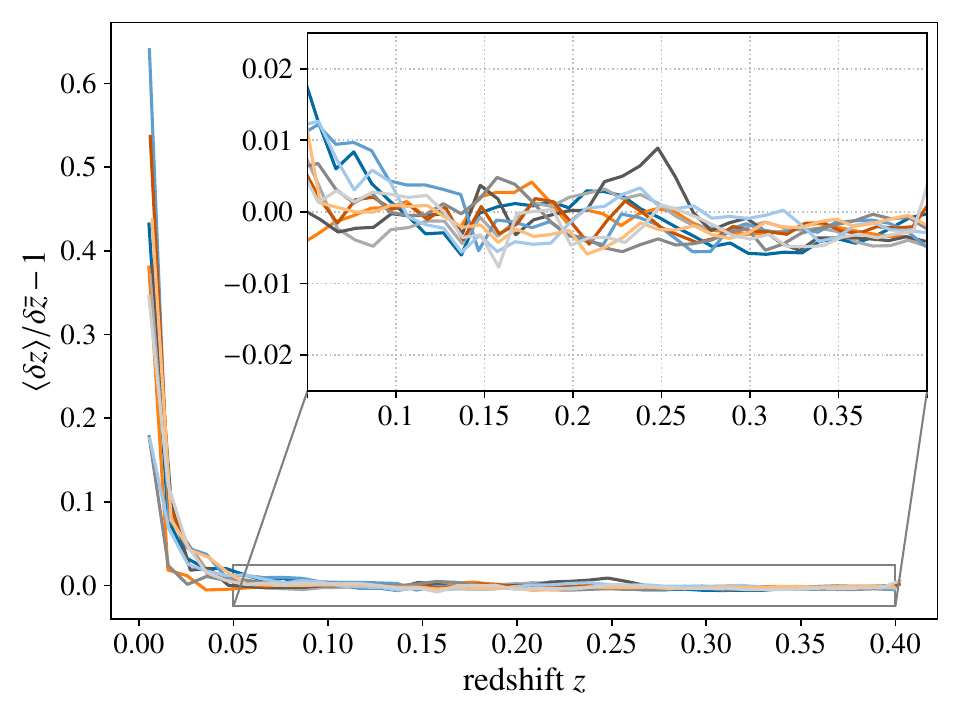}
	\caption{Relative difference between the mean redshift drift in 40 bins and the expected FLRW background result plotted as a function of the redshift for 10 different observers.}
	\label{fig:mean_rsd}
\end{figure}

\begin{figure*}
	\centering
	\includegraphics[width=0.33\textwidth]{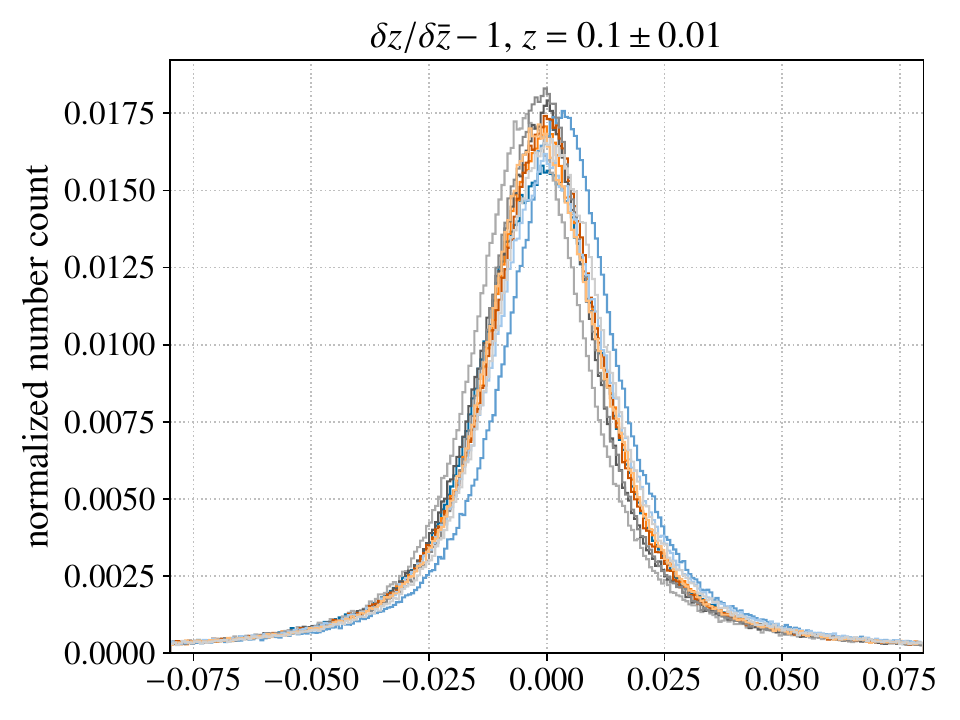}%
	\includegraphics[width=0.33\textwidth]{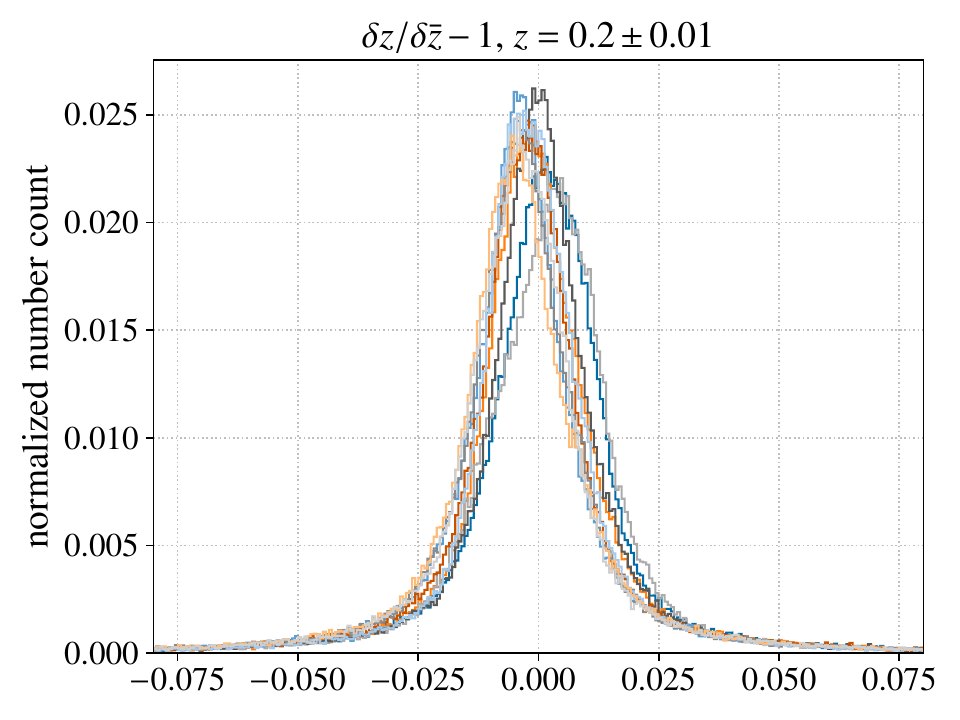}%
	\includegraphics[width=0.33\textwidth]{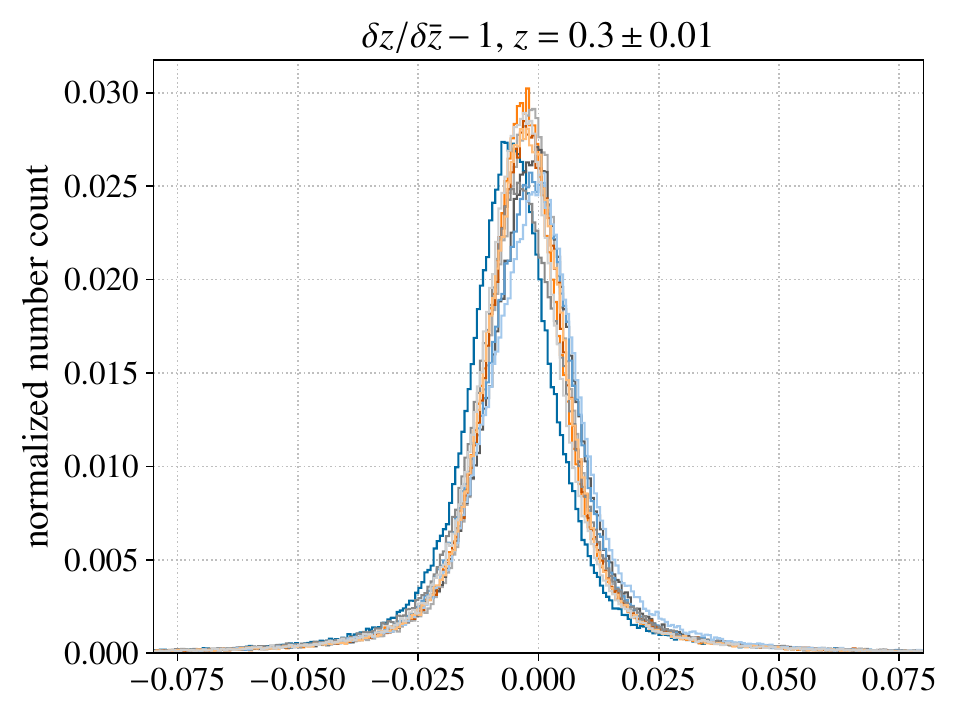}
	\caption{Normalized histograms of the relative fluctuations of the redshift drift around the background FLRW prediction in three different redshift bins $z=0.1,0.2,0.3\pm 0.01$ for 10 different observers.}
	\label{fig:mean_rsd_hist}
\end{figure*}

\begin{figure}
	\includegraphics[width=\linewidth]{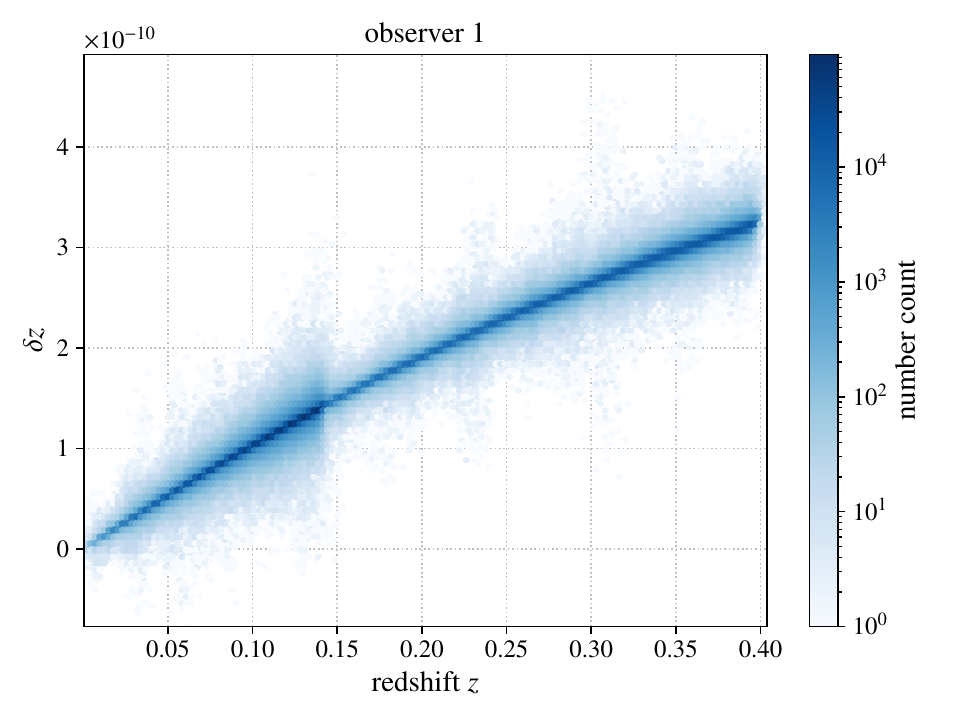}
	\caption{The redshift drift as a function of the redshift for observer 1 plotted as a hexbin density map.}
	\label{fig:rsd_obs1}
\end{figure}

\begin{figure*}
	\includegraphics[width=0.5\linewidth]{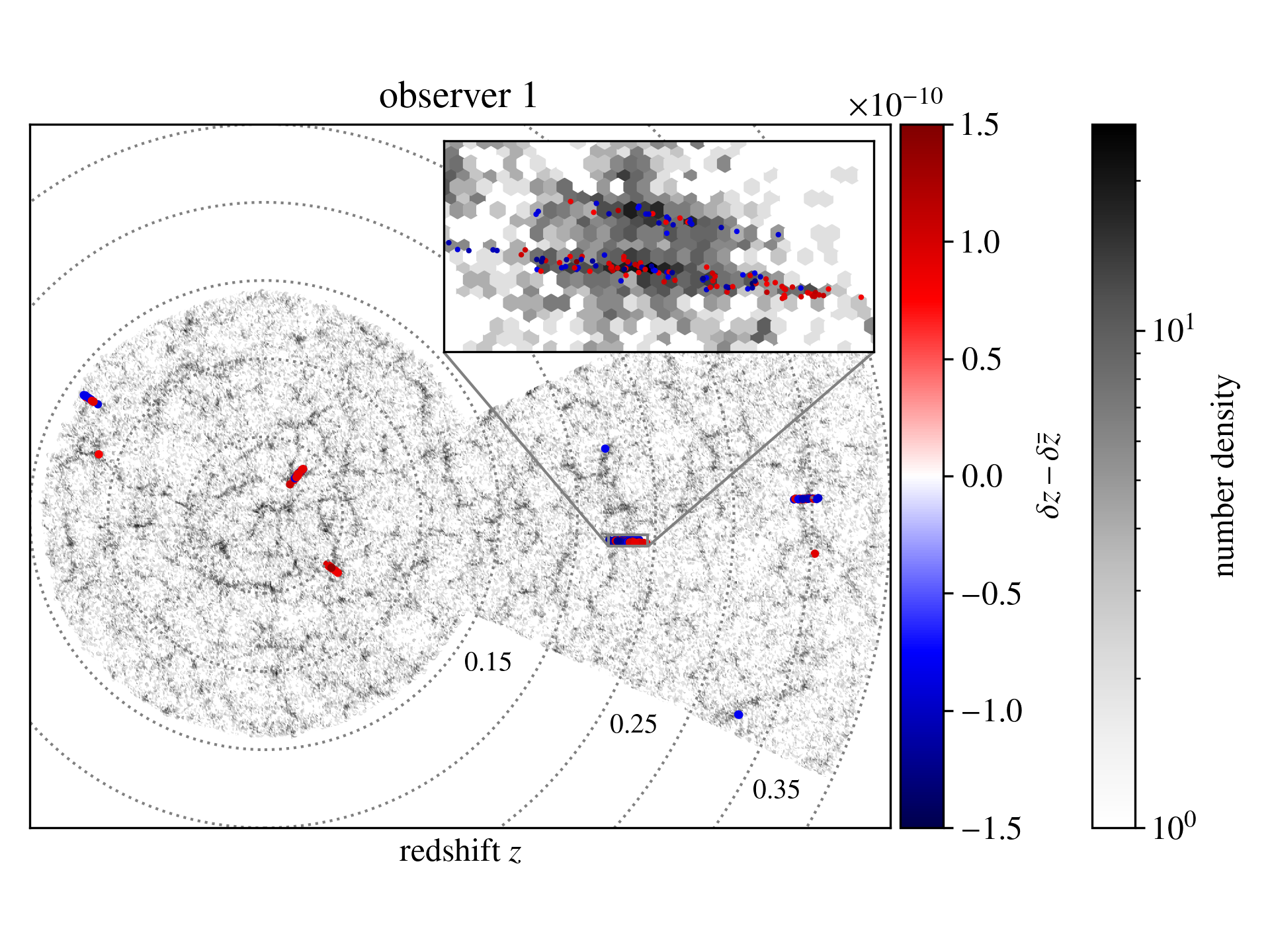}
	\includegraphics[width=0.5\linewidth]{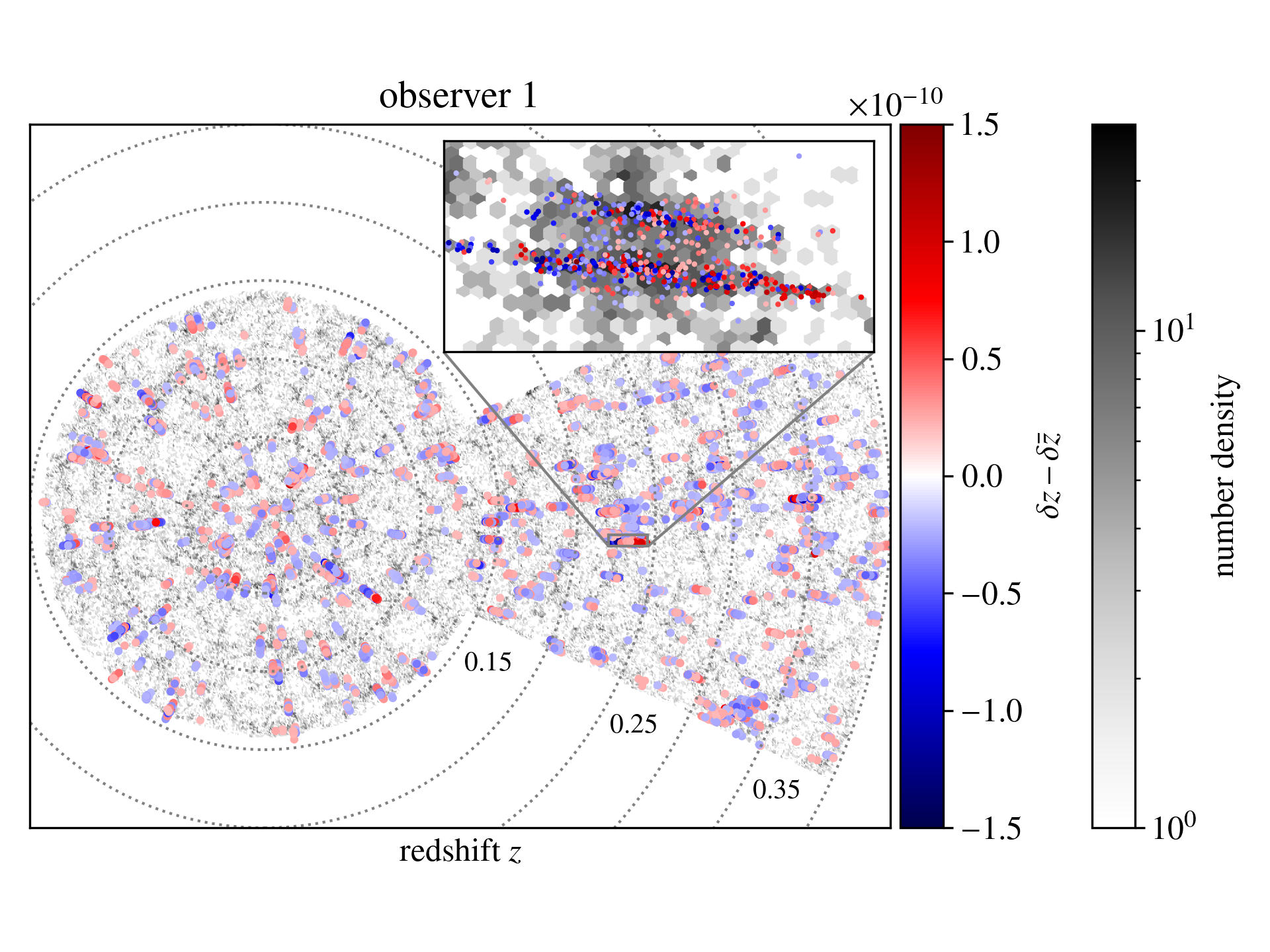}
	\caption{Left: Slice through the observed light cone with width $\d z = 0.01$. Shown are the number density of particles (grey), with those particles highlighted that have $|\delta z -\delta\bar z| > 0.8\times 10^{-10}$. Right: The same as left, but with those particles with $|\delta z -\delta\bar z| > 0.2\times 10^{-10}$ highlighted.}
	\label{fig:cluster_zoom_obs1}
\end{figure*}

\section{Results}
\label{sec:results}
We here show the results obtained from computing the redshift drift in our simulation and compare to the results expected based on the predictions of Sec. \ref{sec:theory}. We computed the redshift drift on the light cone of 10 randomly placed present-time observers. Unless otherwise mentioned, we show results using every one-hundredth particle in the light cone, to keep the data amount manageable. 
\\ \\
We start by showing the mean observed redshift drift compared to the expected FLRW background result for all 10 observers in Fig.~\ref{fig:mean_rsd}. The mean is calculated in 40 bins. At very low redshifts $z<0.05$ we see a large deviation from the background. This is expected given that the very local environment is likely to be either over- or under-dense compared to the mean. For redshifts $0.05<z<0.4$, the mean redshift drift fluctuates around the background within approximately $0.5\%$. At the largest redshifts in our light cone, $0.28\lesssim z \lesssim 0.38$, all observers see a mean redshift drift smaller than the expected background result. While this is surprising and would be very interesting if true in general, 10 observers are not enough to draw a definite conclusion, as it may in principle be due to e.g. special features of the observer position. In addition, the deviation is very small and therefore is not likely to have any implications for real-world redshift drift observations. The difference could potentially be due to the integrated Sachs-Wolf (ISW) effect, i.e. terms that integrate over time derivatives of the Bardeen potentials, arising when considering the contributions of metric fluctuations to the redshift drift, see e.g. \cite{Bessa2023}. To confirm, one would have to investigate if the effect holds for more observers and higher redhifts.
\newline\indent
In Fig.~\ref{fig:mean_rsd_hist} we show normalised histograms of the relative redshift drift w.r.t. the FLRW background expectation for three redshift bins $z=0.1,0.2,0.3\pm0.01$ for all 10 observers. The distribution is approximately Gaussian, with the curves for $z\sim0.3$ being shifted slightly to the negative side, as expected from Fig.~\ref{fig:mean_rsd}. Note that the distribution seems to be only shifted, not skewed. 
\\ \\
Next, we focus on one specific observer. In Fig.~\ref{fig:rsd_obs1} we show the redshift drift as a function of the redshift for observer 1. We show a hexbin density map\footnote{\url{https://matplotlib.org/stable/api/_as_gen/matplotlib.pyplot.hexbin.html}} \cite{matplotlib}, which shows the number of particles in hexagonal bins. One can clearly see the expected mean result, but also some surprisingly large scatter, which is of the same order of magnitude as the background signal at low redshifts. This scatter also exhibits some structure. Although the sharp cut at $z\sim 0.15$ is a result of the transition from the full-sky light cone to the partial cone, there seems to be some large scatter in vertical lines at specific redshifts. These result from particles falling into the cluster potential of galaxy clusters in our line of sight and therefore being highly accelerated. We illustrate this in the left panel of Fig.~\ref{fig:cluster_zoom_obs1}, where we show a thin slice through the observer light cone with thickness $\d z = 0.01$. We plot the number density of particles  and highlight only those particles with $|\delta z -\delta\bar z| > 0.8\times 10^{-10}$. Zooming in on one of the collections of highlighted particles we can see how they lie along our line of sight, in front of and behind a high over-density. The zoomed-in feature lies at roughly redshift $z=0.23$, where we indeed find one of the mentioned vertical lines in Fig.~\ref{fig:rsd_obs1}. Since we only show a thin slice of the light cone, we do not see highlighted particles corresponding to each vertical line. Decreasing the limit to $|\delta z -\delta\bar z| > 0.2\times 10^{-10}$ for highlighted particles in the right panel of Fig.~\ref{fig:cluster_zoom_obs1}, we can see that indeed all of the far outliers lie in over-dense regions/ clusters. We will later see that these particles are the ones with a high peculiar acceleration. 
\\\\
We have shown the results for observer 1 here, but they are representative for all other considered observers, with the only change being where the vertical features related to the clusters appear. 
\\ \\ 
The far outliers in Fig.~\ref{fig:rsd_obs1} offer an intriguing possibility of constraining the peculiar acceleration and possibly reconstructing the potential of galaxy clusters. As can be seen from \eqref{eq:dz_v2} the scatter around the background constrains the Hubble weighted peculiar acceleration field along our line of sight $\d /\d t\, (\bold v / H)\cdot \bold e$. We will see in the next subsection that \eqref{eq:dz_v2} is very accurate. It would be interesting to see how these results could be related to those in \cite{Amendola2008}, where it was first suggested to use the peculiar acceleration to reconstruct the potential of local clusters. To do so, it would be interesting to see how the scatter depends on the mass of the objects studied, e.g. by employing a Halo finder to identify realistic sources.

\begin{figure*}
	\centering
	\includegraphics[width=0.4\textwidth]{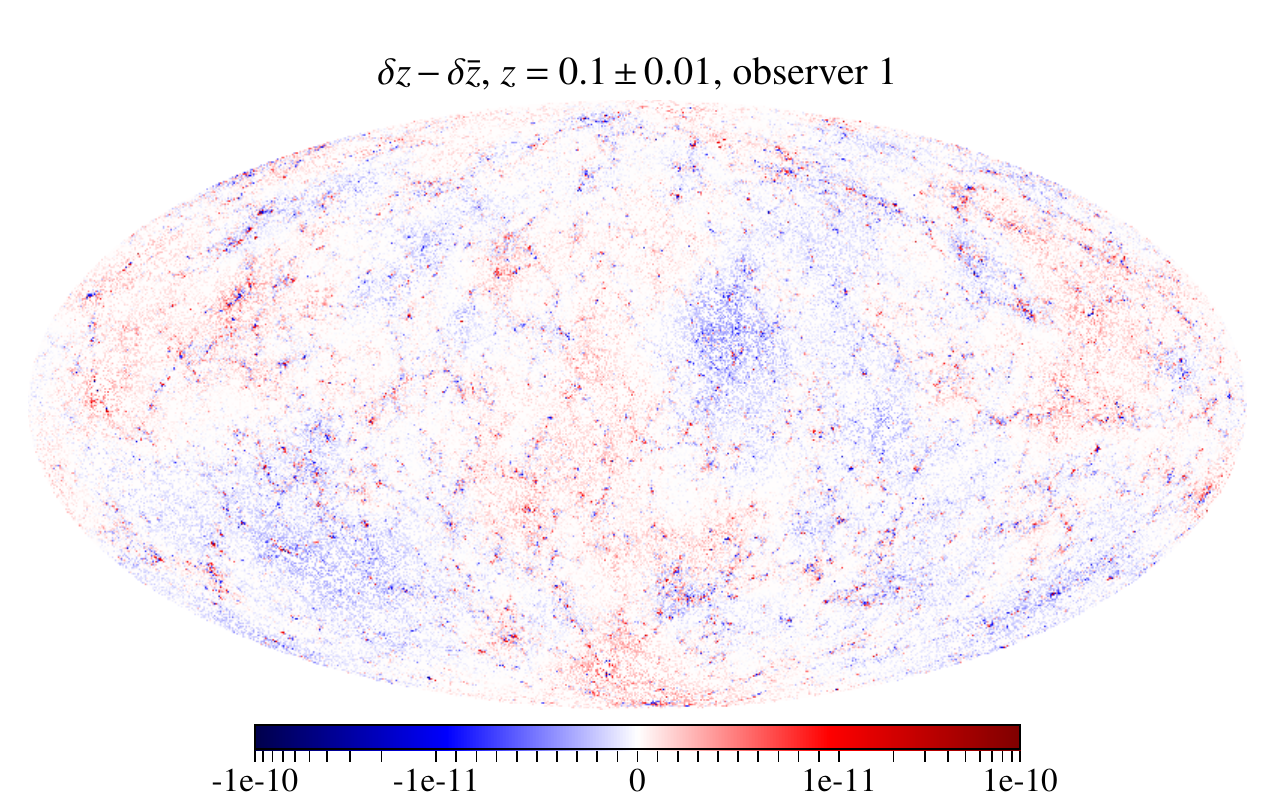}%
	\includegraphics[width=0.24\textwidth]{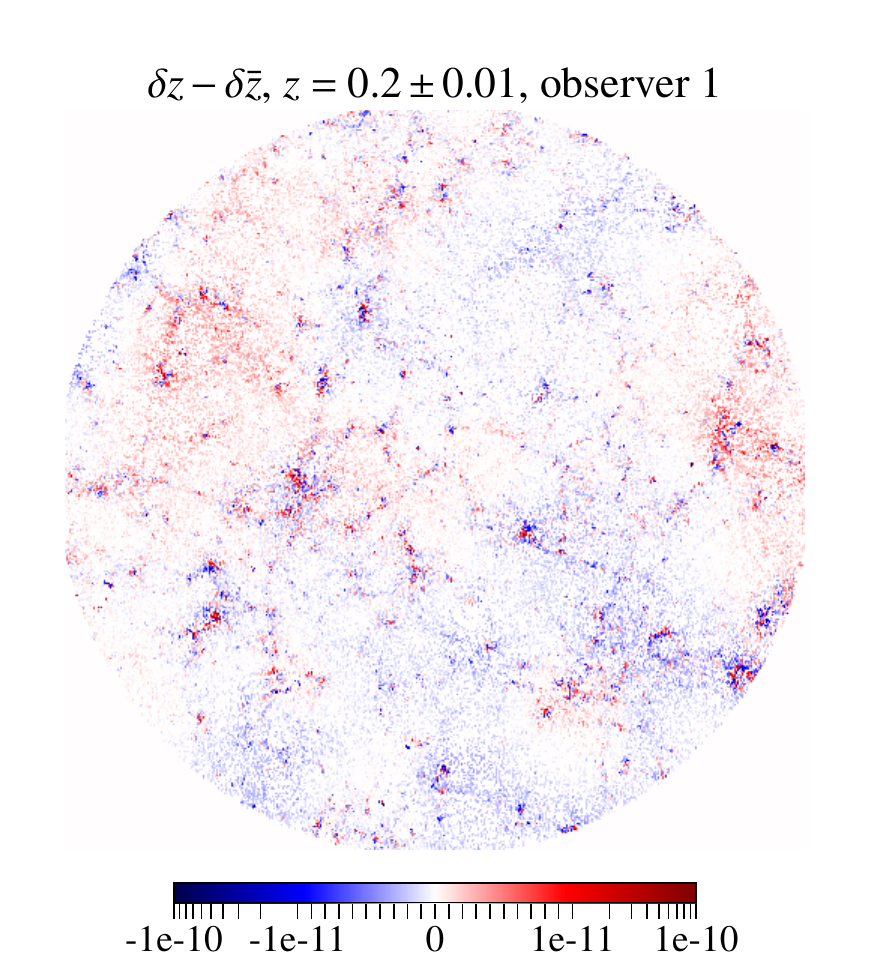}%
	\includegraphics[width=0.24\textwidth]{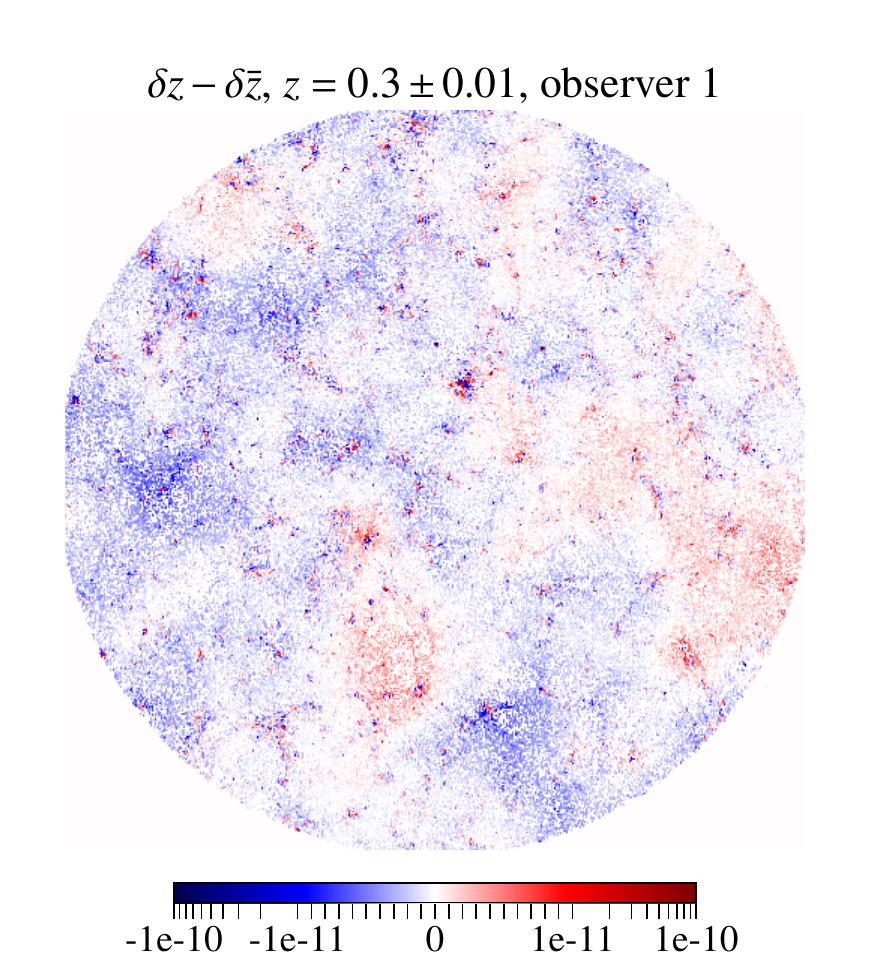}
	\includegraphics[width=0.4\textwidth]{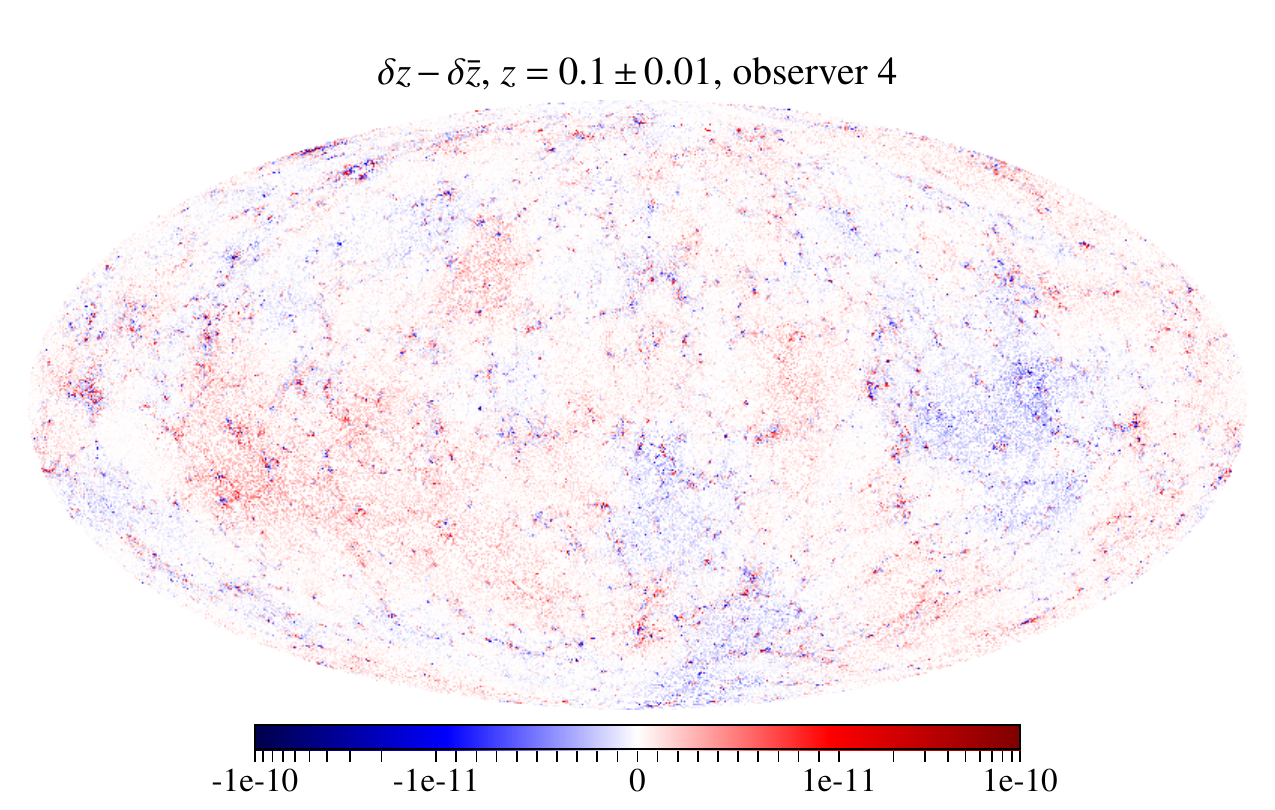}%
	\includegraphics[width=0.24\textwidth]{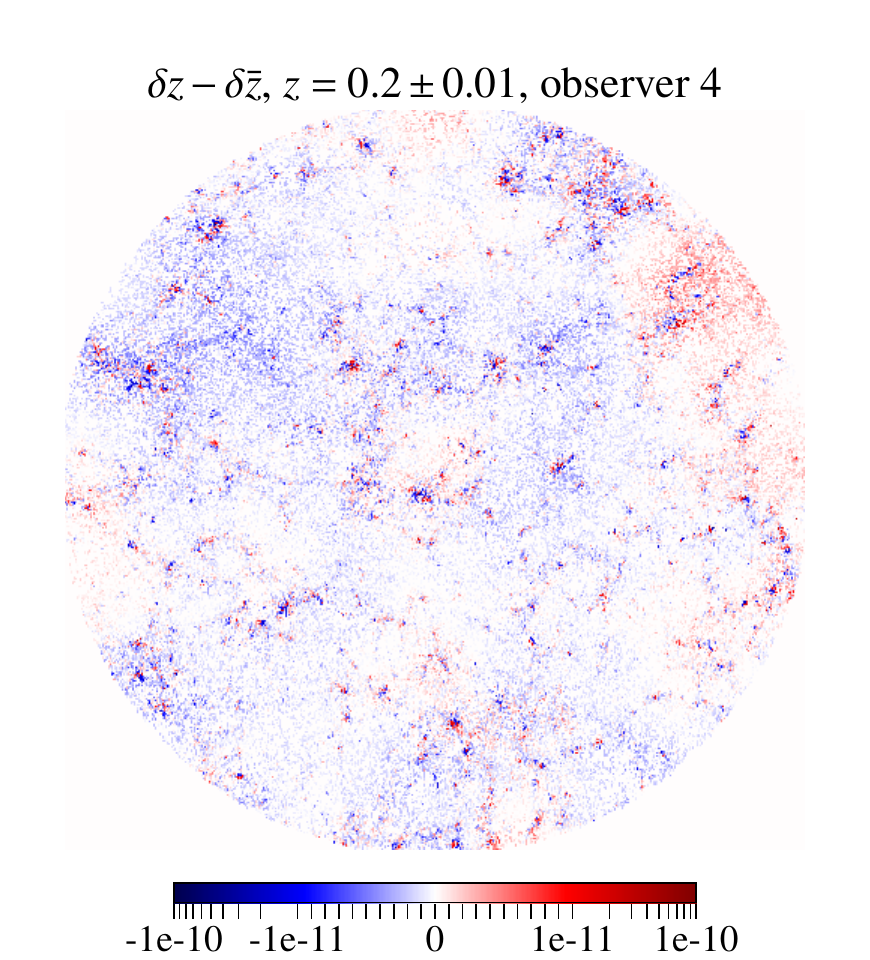}%
	\includegraphics[width=0.24\textwidth]{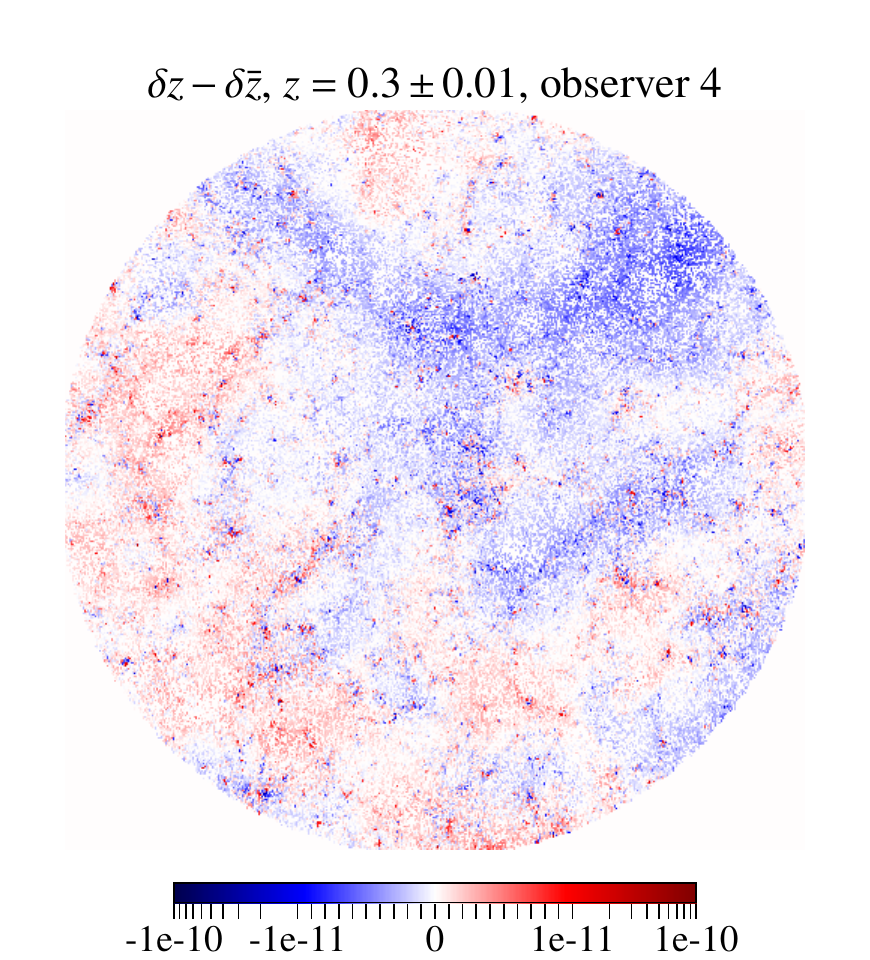}
	\includegraphics[width=0.4\textwidth]{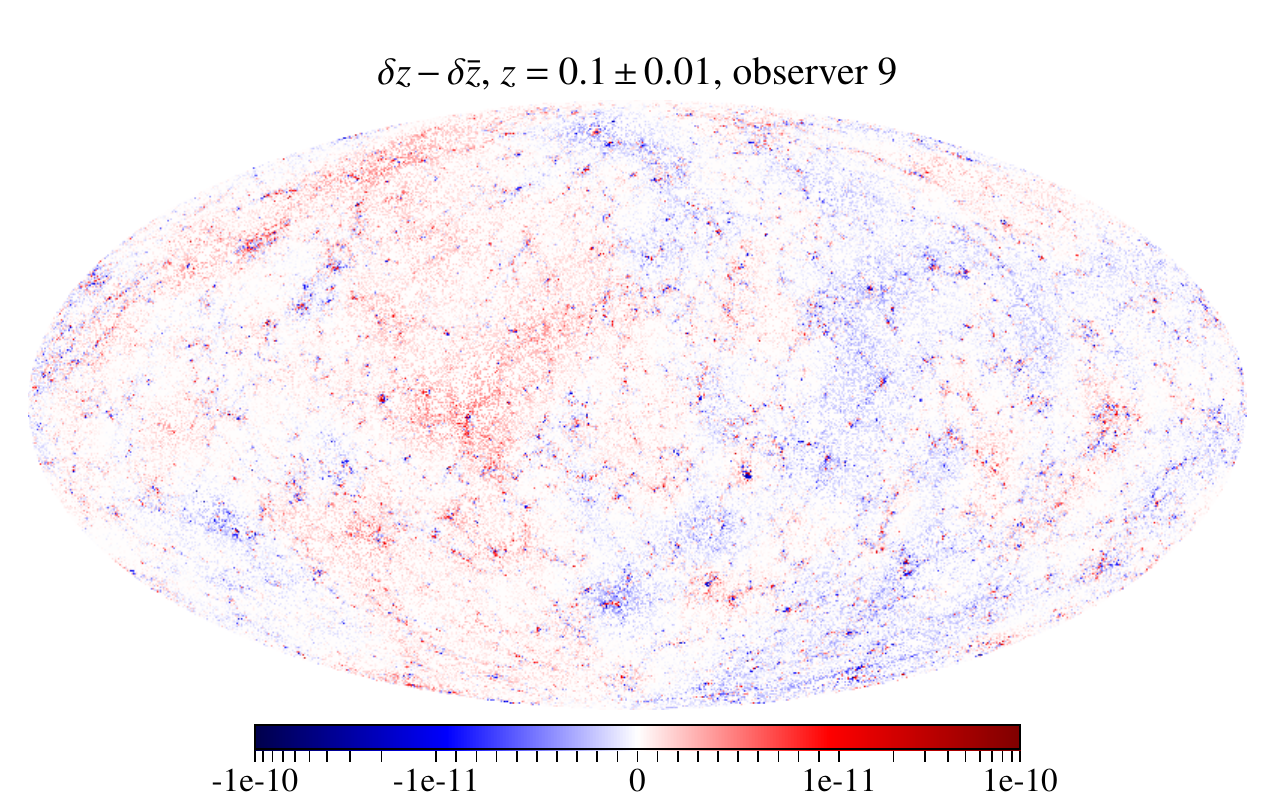}%
	\includegraphics[width=0.24\textwidth]{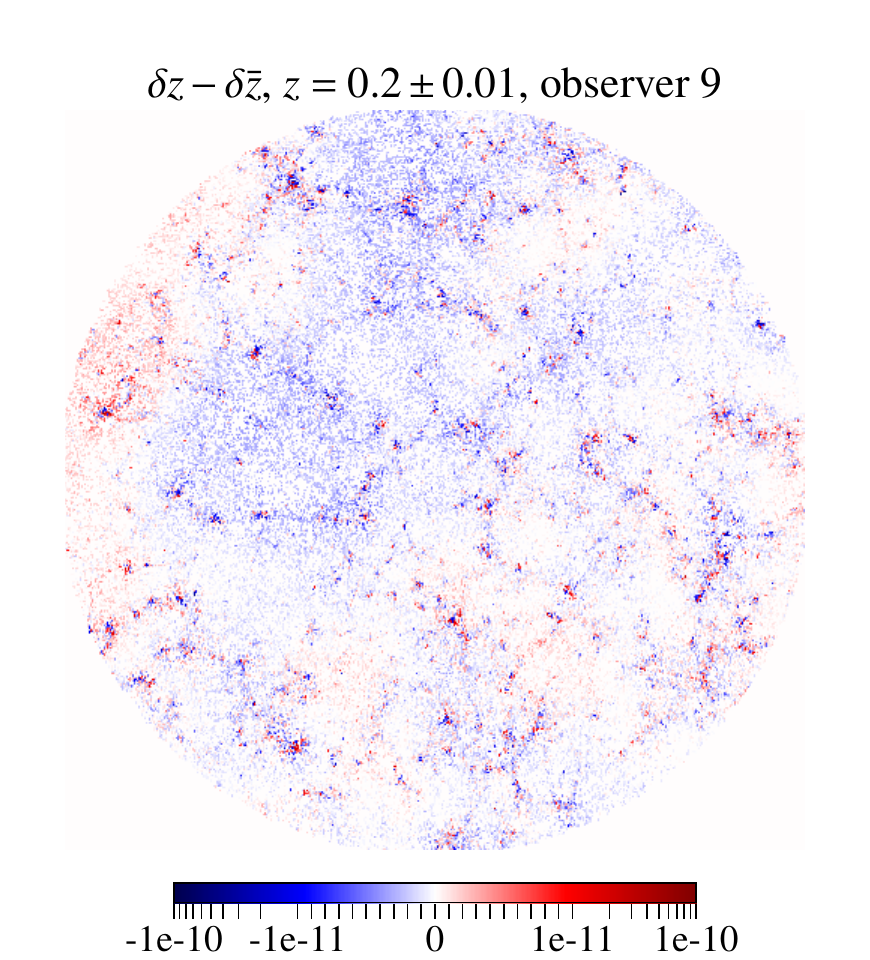}%
	\includegraphics[width=0.24\textwidth]{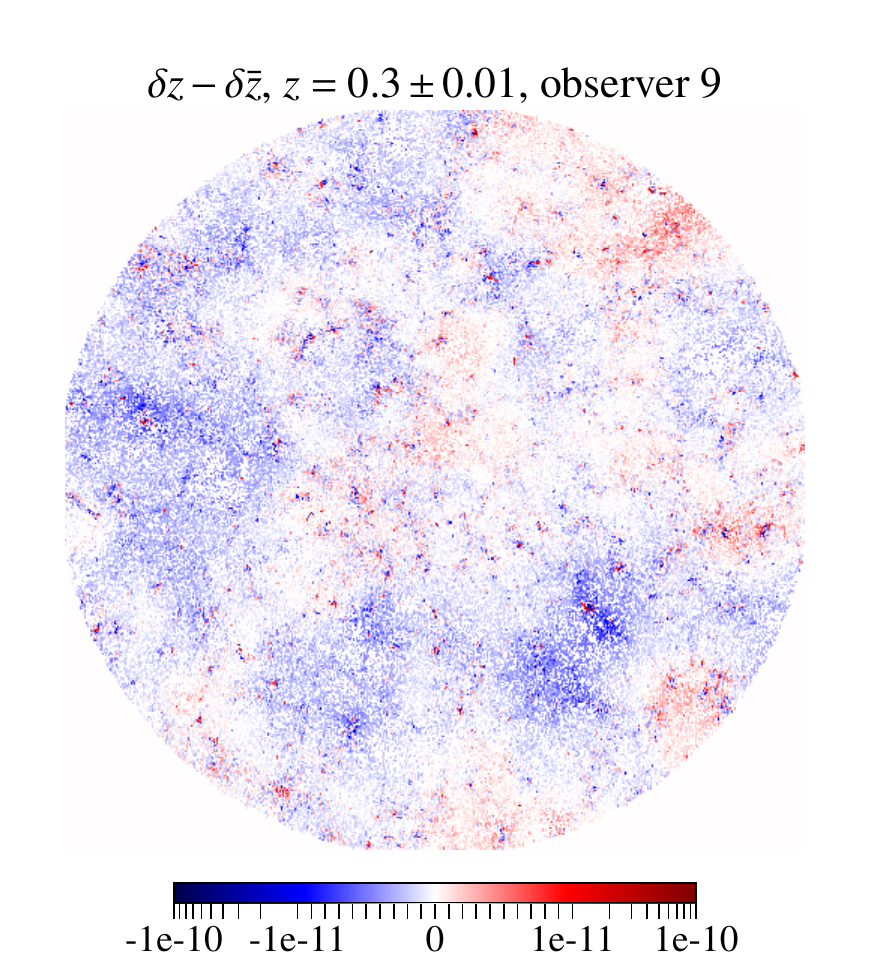}
	\caption{HEALPix skymaps of the redshift drift fluctuations around the FLRW background prediction, for three different redshift bins $z=0.1,0.2,0.3\pm0.01$ and three different observers. The map for $z\sim 0.1$ shows the full sky with resolution $N_\mathrm{side}=256$, while the maps for $z\sim 0.2,0.3$ show partial circular sections of the sky from a pencil beam light cone with half-opening angle 25° and have a resolution $N_\mathrm{side}=512$. These figures are log-scaled between $10^{-10}$ and $10^{-11}$ and linearly scaled between $0$ and $10^{-11}$ to allow both visualization of outliers and the underlying distribution.}
	\label{fig:skymaps}
\end{figure*}

We further note that since the inhomogeneity-induced fluctuations seen in Fig.~\ref{fig:rsd_obs1} are of the same order as the cosmic signal for redshifts up to $z\approx 0.15$, they can lead to a sign change in the redshift drift, albeit only for the more extreme outliers. The possibility of fluctuations to change the sign of the redshift drift is particularly interesting because the sign of the redshift drift in FLRW spacetimes is a direct indication for ($\delta z>0$) or against ($\delta z<0$) accelerated cosmic expansion. The sign of the redshift drift and its relation to accelerated expansion in an inhomogeneous universe have previously been discussed in e.g. \cite{Koksbang2019, Koksbang2020, Heinesen2021b, Koksbang2024} with the conclusion that the sign is still determined by cosmic acceleration, at least for `cosmological redshifts', i.e. with the possible exception of low redshifts where peculiar motion can dominate the signal. Our results support this overall conclusion, although we note that this is the first time a change in the sign of $\delta z$ due to inhomogeneities at redshifts up to $z\approx 0.15$ has been reported. We also note that we will show in the following that the extreme outliers and hence change in the sign of the redshift drift are due to extreme peculiar acceleration. We thus still find that it is the acceleration that determines the sign of the redshift drift, but it is the total local acceleration between emission and observation, not the cosmic acceleration independently.
\\ \\ 
In Fig.~\ref{fig:skymaps} we show HEALPix sky maps of the redshift drift fluctuations at redshifts $z=0.1,0.2,0.3\pm0.01$ for 3 randomly selected observers. We choose $N_\mathrm{side}=256$ for the allsky map at $z\sim 0.1$ and $N_\mathrm{side}=512$ for the pencil-beam maps at $z\sim 0.2,0.3$. If there are multiple particles in one cell of the map, we average their redshift drift. In order to be able to see both the outliers and the underlying distribution, the maps are log-scaled between $10^{-10}$ and $10^{-11}$ and linearly scaled between $0$ and $10^{-11}$. The distribution follows the large scale structure and we again find the far outliers in regions of high density/ many particles. The maps for $z=0.3$ have more blue than red points due to the already mentioned shift in the mean with respect to the expected FLRW result.
\\ \\ 

\subsection{Comparison to Analytical Predictions}
\label{sec:results_theor_pred}
In order to compare our results to the predictions from Sec.~\ref{sec:theory}, we now calculate the redshift drift from equation \eqref{eq:dz_v1}. We approximate the velocity and acceleration of the particles by using the values of the velocity recorded on the two background light cones output by the simulation. For the velocity we use the value from the $z=0$ light cone and for the acceleration we simply take the finite difference quotient of the values on the two light cones
\begin{equation}
    \dot{\bold v}_\s = \frac{\d \bold v_\s}{\d t_\s} = \frac{\d \bold v_\s}{\d t_\o}\frac{\d t_\o}{\d t_\s} \approx \frac{\bold v_\s(\tau_1)-\bold v_\s (\tau_2)}{t_\o(\tau_1)-t_\o(\tau_2)}(1+z)\;,
\end{equation}
where it is important to account for the fact that we want $\dot{\bold v}_\s$ at the source position, but only have access to $\d t_\o$ at the observer position. We also need the direction vector $\bold e$ which can be calculated directly from the observed position on the sky recorded by the raytracer. Lastly, a factor $1/c$ has to be restored to the velocity and acceleration term to fix the units. 

\begin{figure*}
	\centering
	\includegraphics[width=0.33\linewidth]{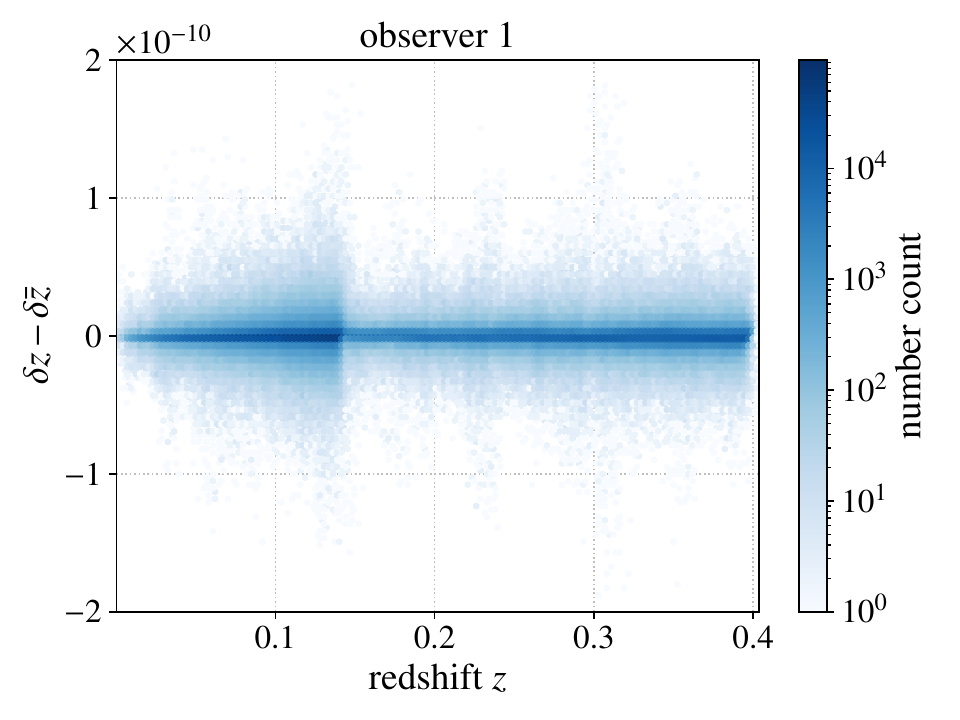}%
	\includegraphics[width=0.33\linewidth]{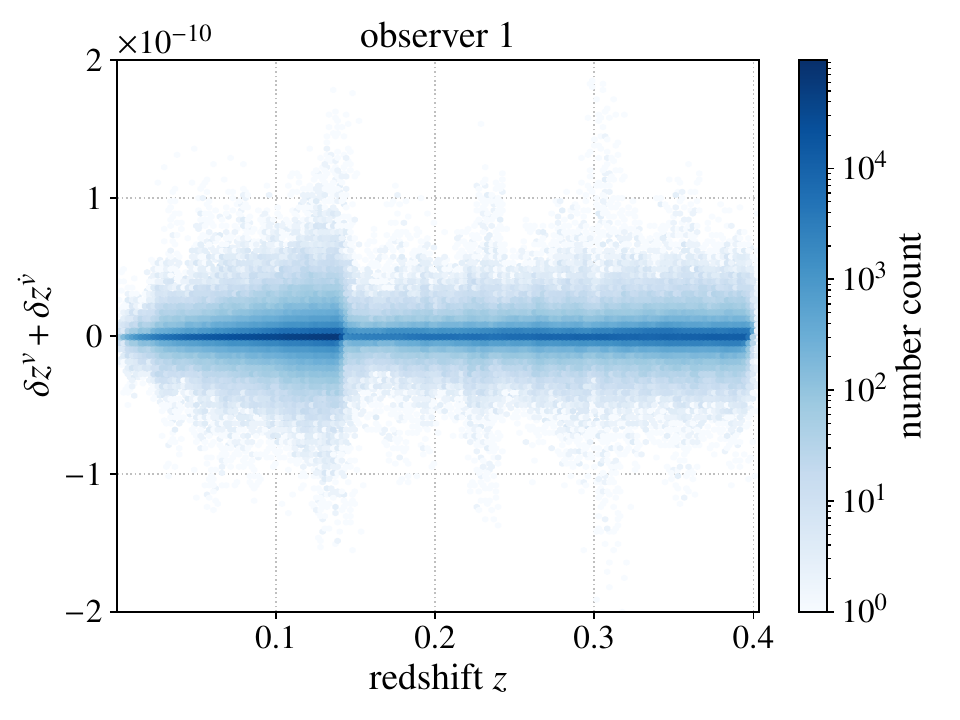}%
	\includegraphics[width=0.33\linewidth]{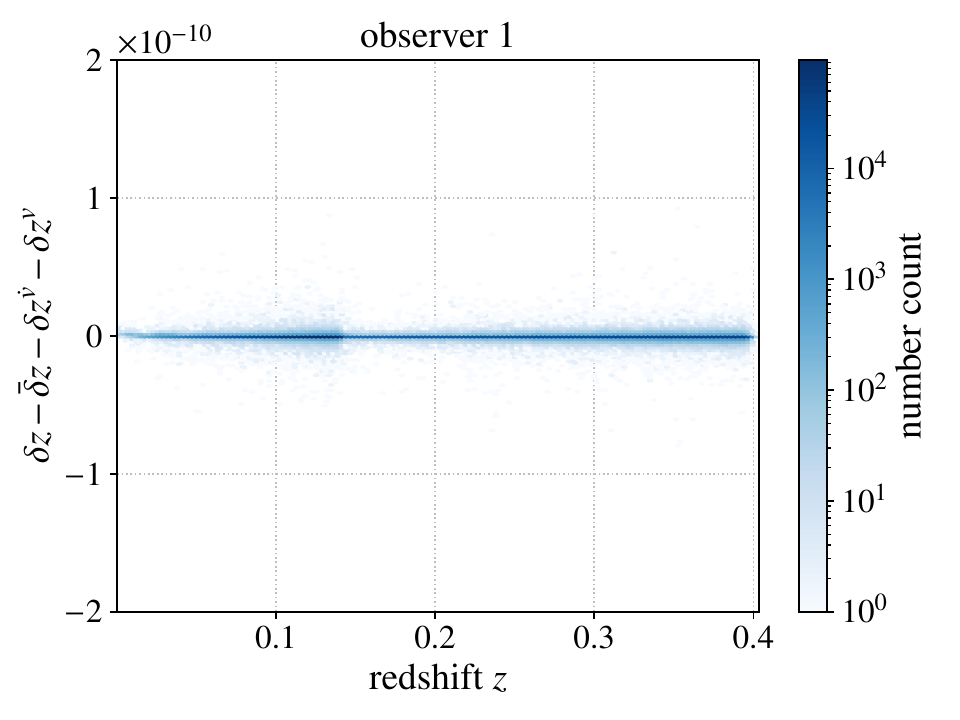}
	\includegraphics[width=0.33\linewidth]{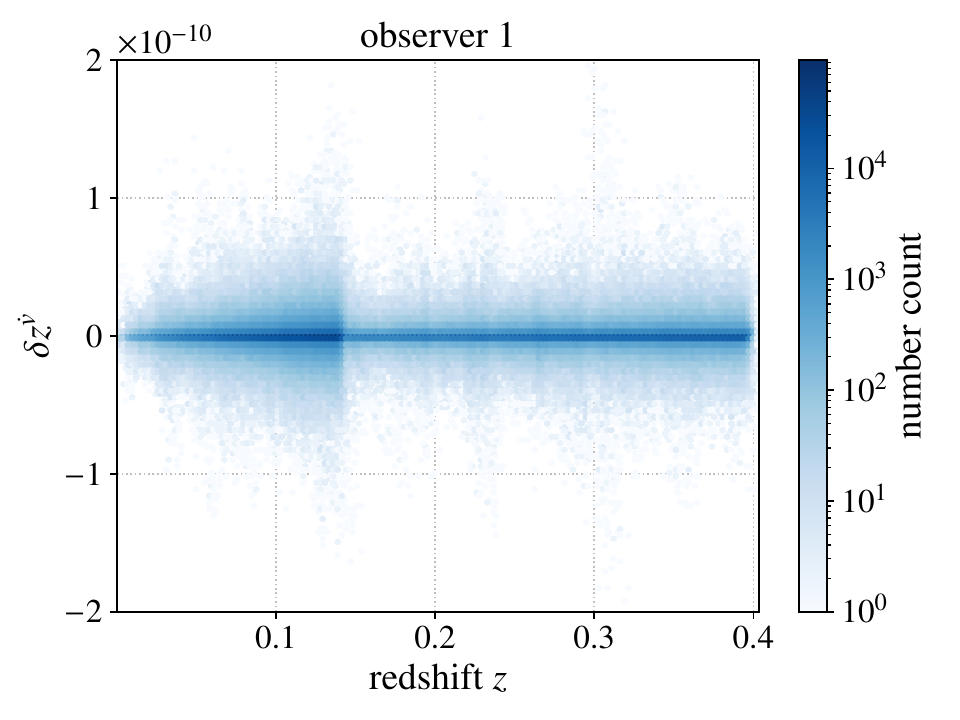}%
	\includegraphics[width=0.33\linewidth]{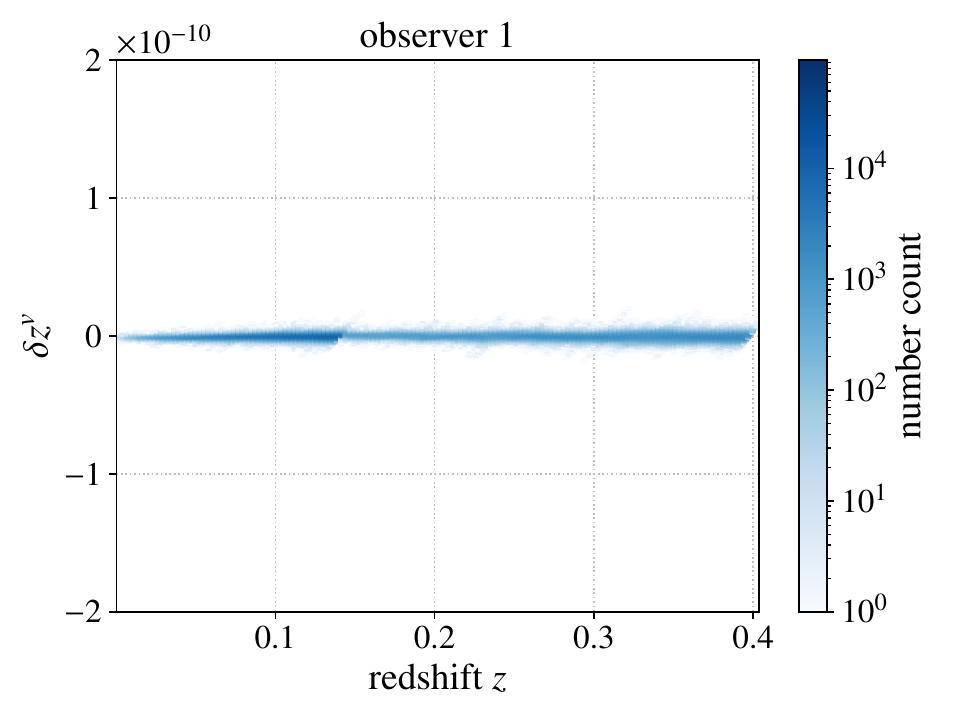}
	\caption{Fluctuations of the redshift drift as functions of the redshift, plotted as hexbin density maps. Simulation result (top left), predictions from perturbation theory according to \eqref{eq:dz_v1} (center top), difference between simulation and predicted result (top right), as well as the individual contributions by the peculiar acceleration (bottom left) and velocity (bottom right) terms as defined in \eqref{eq:peculiar_vel_acc}.}
	\label{fig:pertb_theory_density_maps}
\end{figure*}

\begin{figure*}
	\centering
	\includegraphics[width=0.33\linewidth]{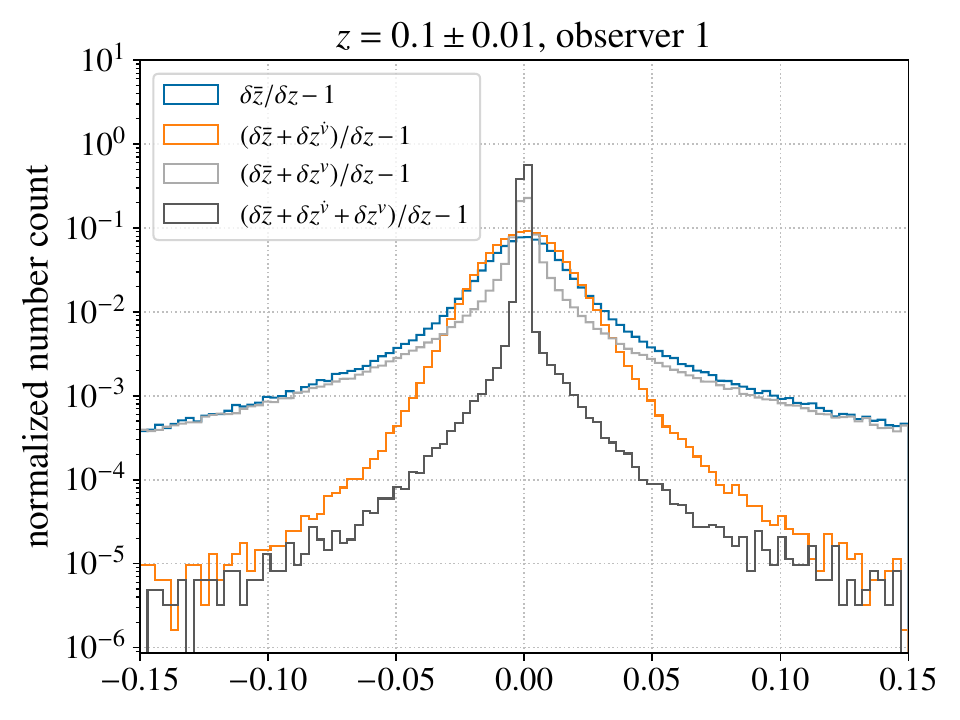}%
	\includegraphics[width=0.33\linewidth]{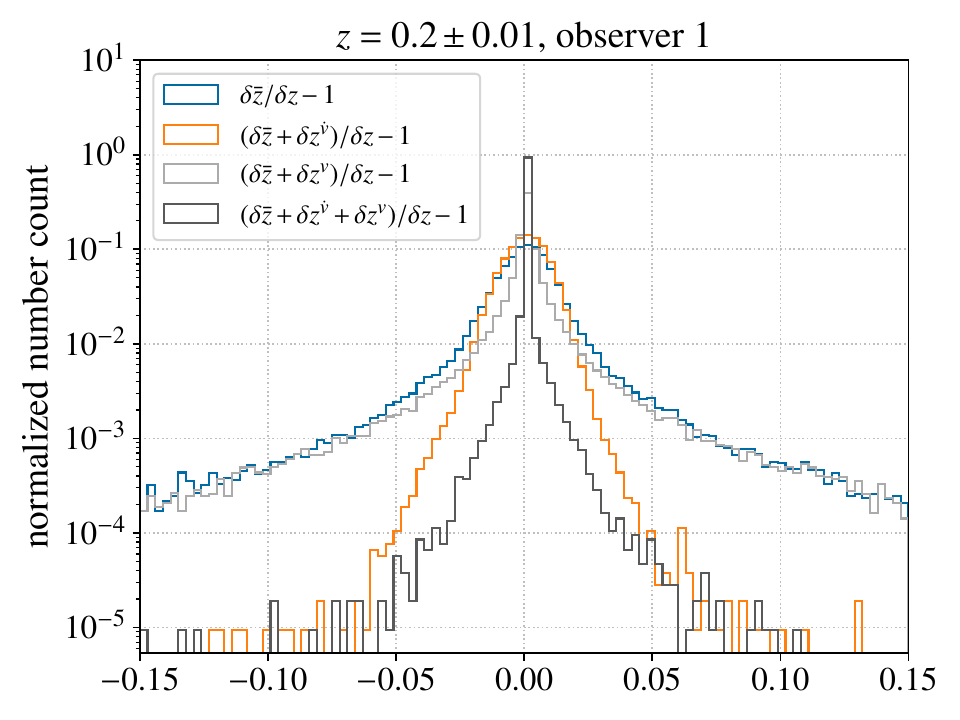}%
	\includegraphics[width=0.33\linewidth]{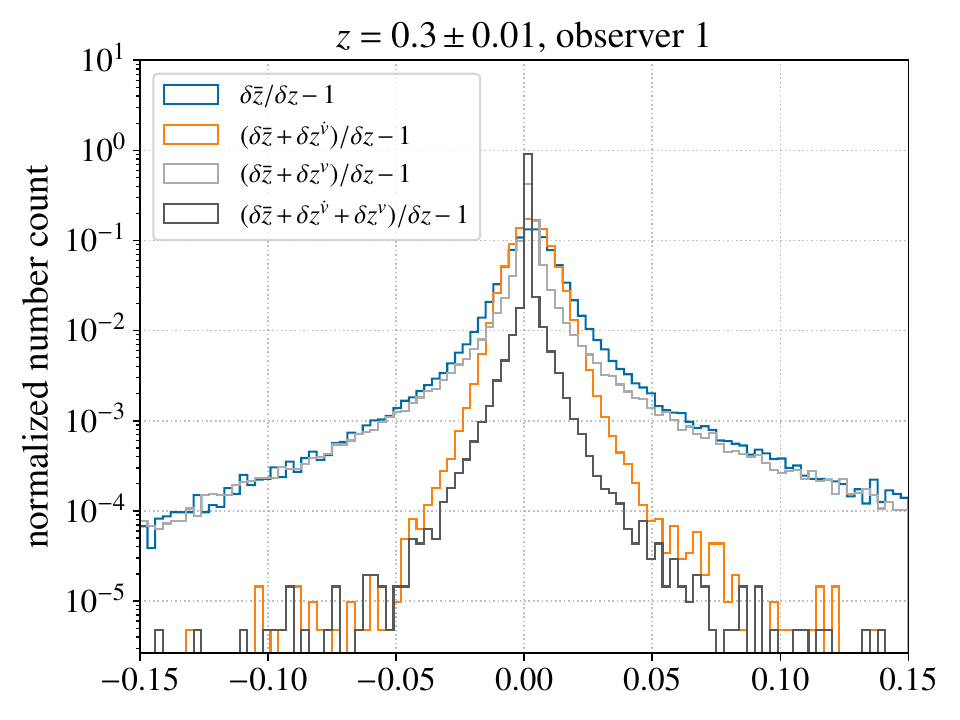}
	\caption{Normalized histograms for the relative difference between the redshift drift and the predictions by perturbation theory according to \eqref{eq:dz_v2} (black). As well as the relative difference accounting only for some of the terms in \eqref{eq:dz_v2}, background (blue), background and peculiar acceleration (orange), and background and peculiar velocity (grey). Shown for three different redshift bins $z=0.1,0.2,0.3\pm 0.01$.}
	\label{fig:pertb_theory_hist}
\end{figure*}

The results are shown in Figs.~\ref{fig:pertb_theory_density_maps},\ref{fig:pertb_theory_hist}. In Fig.~\ref{fig:pertb_theory_density_maps} we again show hexbin density maps, now depicting the redshift drift fluctuations and the individual contributions predicted by \eqref{eq:dz_v1}. The top row shows the redshift drift fluctuations $\delta z-\delta \bar z$ in the leftmost panel and the analytical expectation $\delta z^v+\delta z^{\dot v}$ in the middle panel; by eye, the two seem to be an almost perfect match. Indeed, when subtracting all contributions from $\delta z$ (top row leftmost panel), we can see that we almost perfectly remove all the scatter. In the bottom row, we show the individual contributions by the acceleration (left) and velocity (right). We see that the peculiar acceleration is responsible for all the far outliers, while the peculiar velocity produces a much smaller scatter. We have seen earlier in Fig.~\ref{fig:cluster_zoom_obs1} that all of the far outliers are sitting in over-densities/ cluster. With Fig.~\ref{fig:pertb_theory_density_maps} we have now confirmed that these are highly accelerated particles that fall into the cluster potentials.

To better quantify how well the predictions from Sect.~\ref{sec:theory} work, we show normalised histograms for three different redshifts in Fig.~\ref{fig:pertb_theory_hist}. We plot the relative difference between the prediction of the analytical  and the simulated results for $\delta z$ as well as the relative difference, taking only some of the analytically derived terms into account. We can see that, when taking all terms into account, Eq.\ \eqref{eq:dz_v1} predicts the redshift drift to better than one percent accuracy for almost all particles. If we include only the background and peculiar velocity contributions we get a very narrow distribution close to zero with wide tails, i.e.\ this simplified theoretical prediction is good for particles with small scatter. If we instead include the background and peculiar acceleration term, we find a wide distribution around zero that rapidly falls off, i.e.\ we are now predicting particles with large scatter well. This confirms our earlier observations. 

These results show that \eqref{eq:dz_v1} predicts the observed redshift drift remarkably well. Contributions from metric fluctuations and higher-order velocity terms can therefore safely be neglected.

\begin{figure*}
	\centering
	\includegraphics[width=0.33\linewidth]{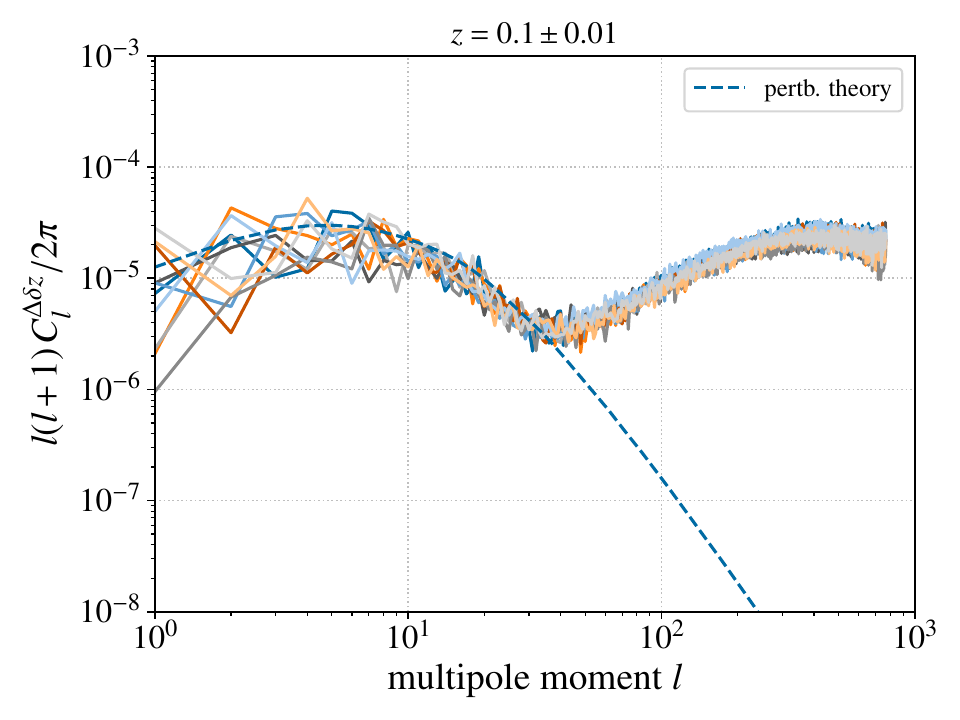}%
	\includegraphics[width=0.33\linewidth]{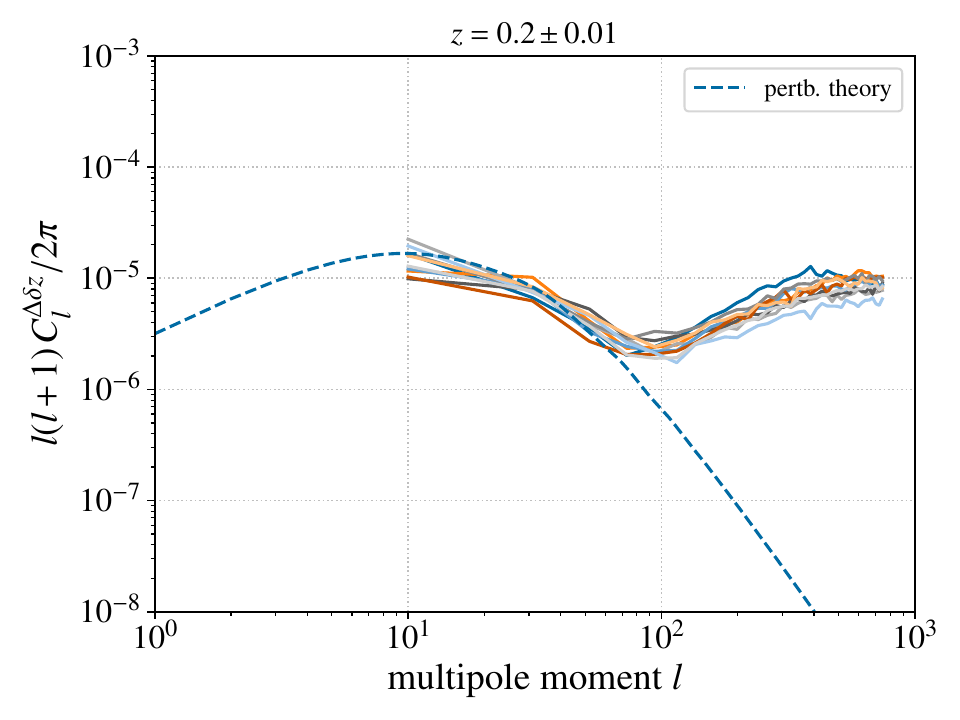}%
	\includegraphics[width=0.33\linewidth]{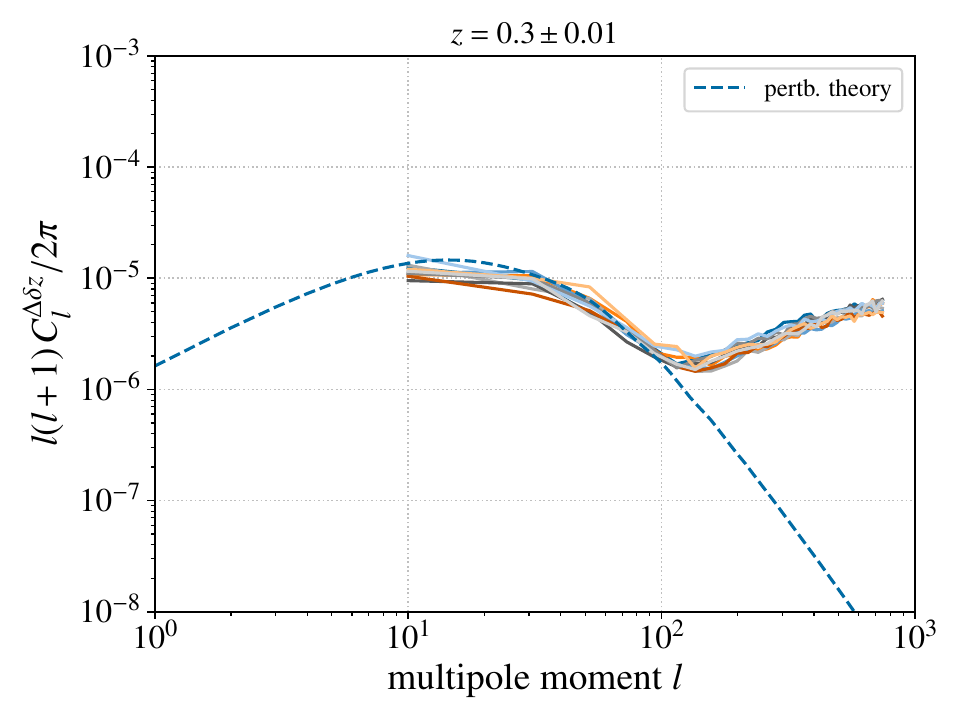}
	\caption{Angular power spectra for the redshift drift fluctuations for 10 different observers compared to the predictions by linear perturbation theory according to \eqref{eq:rsd_Cl}. Results are shown for three different redshift bins $z=0.1,0.2,0.3\pm 0.01$ (left, center, right). The spectra are calculated from HEALPix maps with side length $N_\mathrm{side}=256$.}
	\label{fig:rsd_cl}
\end{figure*}

\subsection{Angular Power Spectra}
\label{sec:power_spectra}
Since the raytracer outputs not only the redshift $z$ for each particle, but also the observed position on the sky in HEALPix coordinates, it is straightforward to calculate angular power spectra for the relative redshift drift fluctuations 
\begin{equation}
	\Delta\delta z = \frac{\delta z}{\delta\bar z}-1\;.
\end{equation}
For redshifts $z=0.1,0.2,0.3$ we sort those particles with redshift $z\pm 0.01$ into HEALPix maps with $N_\mathrm{side}=256$, recording the average value of $\Delta\delta z$, if there are multiple particles in a cell. We now use every tenth particle in the light cone in order to have sufficient statistics and not end up with any empty cells on the HEALPix map. In order to reduce shot noise we create two maps per redshift with half of the particles each, with particles being split randomly into the two maps. We then estimate the auto-correlation spectrum by calculating the cross-spectrum between the two maps. This is known as jackknife resampling (see e.g. \cite{Lepori2021}). With these maps, we now calculate the spectrum using HEALPix's function \texttt{anafast}\footnote{\url{https://healpix.sourceforge.io/html/fac_anafast.htm}}. For the two maps at $z=0.2,0.3$, which only have partial sky coverage, we pass a masked map, masking the empty sky regions, divide the spectrum by the sky-fraction $C_l/f_\mathrm{sky}$ to correct the amplitude, and bin the spectra with $\Delta l \approx 1/f_\mathrm{sky}$ in order to reduce statistical noise. For binning the spectra we use the python script \texttt{bin\_llcl.py} included with the code \texttt{PolSpice}\footnote{\url{https://www2.iap.fr/users/hivon/software/PolSpice}} \cite{Chon2004,Szapudi2001}. As a cross-check, we also calculated the power spectra using \texttt{PolSpice}, confirming that results we obtained with \texttt{anafast} and \texttt{PolSpice} agree.
\\ \\
The resulting power spectra can be seen in Fig.~\ref{fig:rsd_cl}. For very small multipole moments $l$ the spectra follow the prediction from linear perturbation theory very well. Depending on the redshift we see a surprisingly sharp increase in power somewhere between $3<l<20$. This increase is due to the non-linear signal from the peculiar velocities and acceleration. We ensure the signal is not shot-noise or other type of numerical noise in Appx.~\ref{appendix:ps_validation}. Such a strong signal from non-linearities for such small $l$ is surprising and has not been seen in earlier simulations of the redshift drift power spectrum. 

\begin{figure*}
	\centering
	\includegraphics[width=0.33\linewidth]{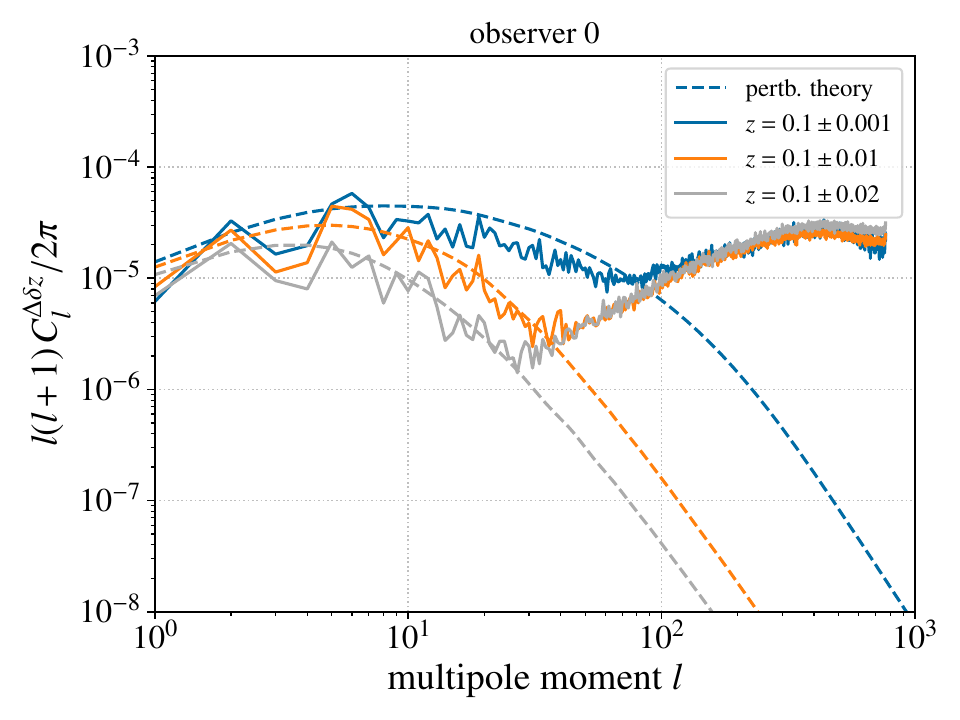}%
	\includegraphics[width=0.33\linewidth]{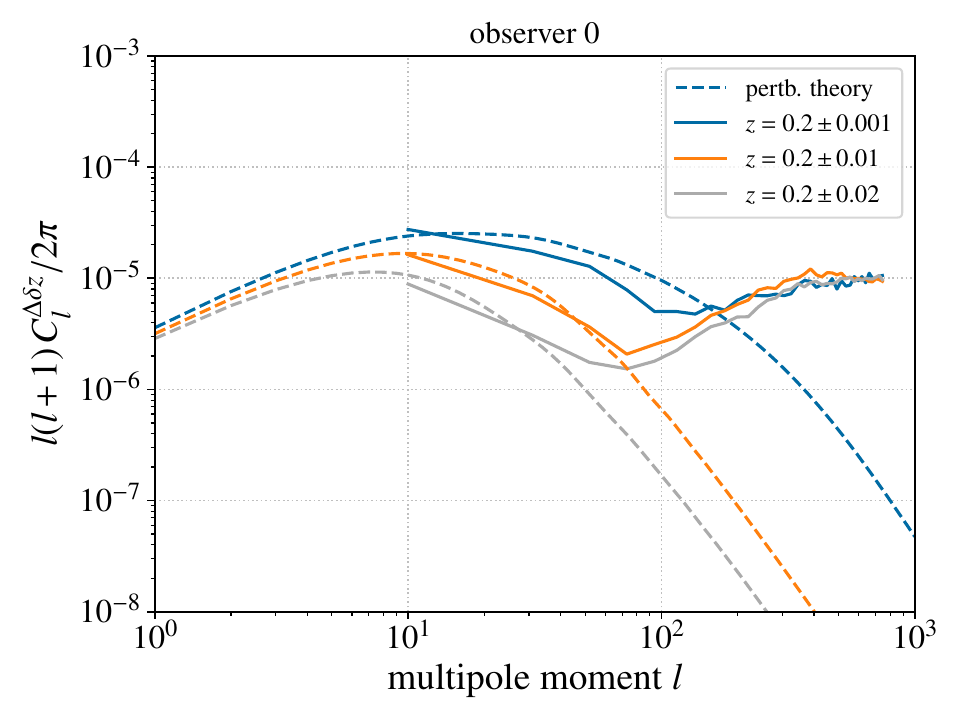}%
	\includegraphics[width=0.33\linewidth]{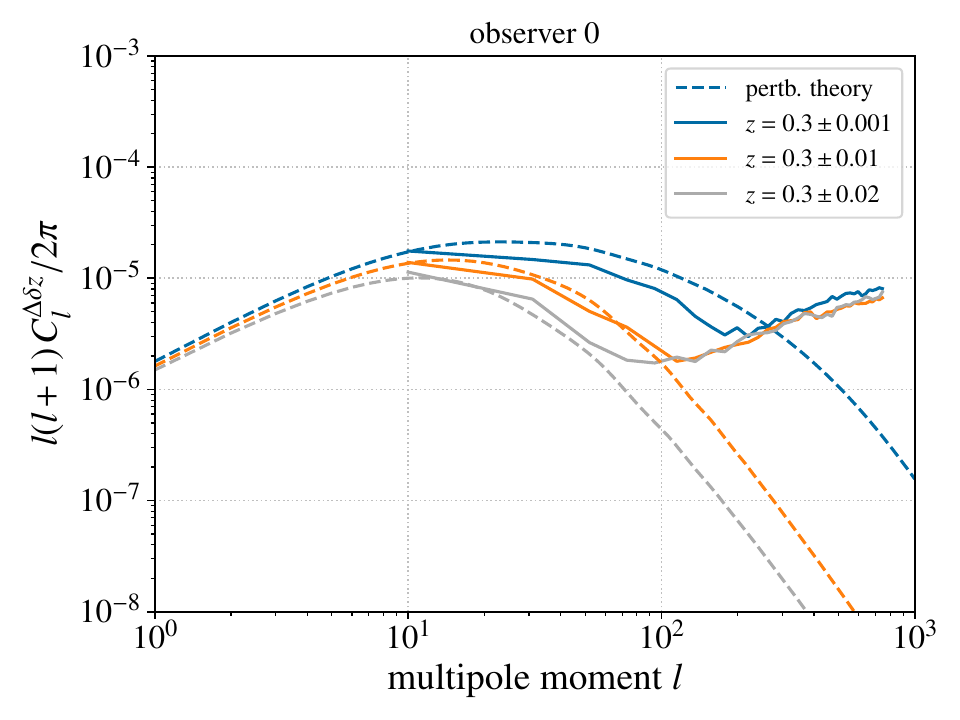}
	\caption{Angular power spectra for the redshift drift fluctuations for  a single observer compared to the predictions by linear perturbation theory according to \eqref{eq:rsd_Cl}. Results are shown for three different redshift bins $z=0.1,0.2,0.3$ (left, center, right), each with varying bin size $\d z= 0.001,0.1,0.2$. The spectra are calculated from HEALPix maps with side length $N_\mathrm{side}=256$.}
	\label{fig:rsd_cl_dz}
\end{figure*}

As a side-note, we note that the window function has significant effects on the power spectra. This is demonstrated in Fig.~\ref{fig:rsd_cl_dz}, where we show the power spectrum for different widths of our window function $\d z=0.001,0.1,0.2$. Here, we use all particles in the light cone in order to have sufficient particles to fill the sky map with the narrowest window function, and do this for one observer only due to the numerical cost. One can see that the larger the Tophat, the earlier (smaller $l$) the non-linear signal can be seen. For larger $l$ the spectra seem to somewhat converge independent of the size of the window function.
\\ \\
Power spectra for the redshift drift fluctuations were also calculated from simulations in \cite{Koksbang2024}, but included no comparison to predictions from perturbation theory. Using our result from Sect.~\ref{sec:theory_ps} we now make this comparison. In \cite{Koksbang2024}, the authors used a fully general relativistic simulation of structure formation in an Einsten-de Sitter (EdS) universe made with the Einstein Toolkit \cite{Macpherson2017, Macpherson2019}. The initial conditions were set using CLASS, with $h=0.45$ and otherwise standard CLASS parameters. We generate an initial spectrum in the same manner and use the EdS linear growth factor to evolve it to present time for use in \eqref{eq:rsd_Cl}.  The authors of \cite{Koksbang2024} calculated the power spectrum from HEALPix maps with $N_\mathrm{side}=32$ for two different redshifts $z=0.1,0.5$ and 10 observers each. Their raytracing algorithm traces to the redshift closest to the two target redshifts. We plot the distribution of redshifts in the individual map pixels for each observer and find that the resulting histogram is best described by a Gaussian with mean $\langle z\rangle = 0.108,0.508$ and width $\sigma_z=0.002, 0.0025$ respectively. We therefore use such Gaussian's as our window function for calculating the linear prediction of the power spectrum. Due to the low number of pixels in the HEALPix maps, this is an approximation, and we cannot expect a perfect match with the simulation results. 

\begin{figure}
    \centering
    \includegraphics[width=0.85\linewidth]{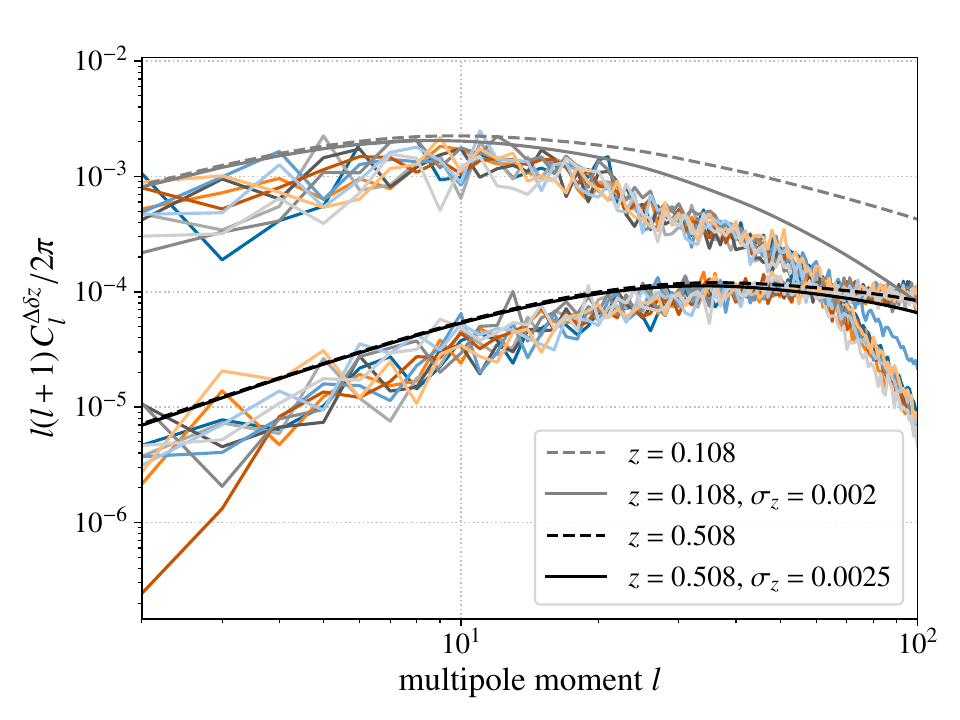}
    \caption{Comparison between predictions for the angular power spectrum of redshift drift fluctuations from perturbation theory as calculated in this paper and simulation results by \cite{Koksbang2024}.}
    \label{fig:ps_comparison_koksbang24}
\end{figure}

The results can be seen in Fig.~\ref{fig:ps_comparison_koksbang24} where we show both the prediction using the Gaussian window function (solid grey and black lines) as well as a delta function (dashed grey and black lines) to demonstrate how important the window function is, especially for redshift $z=0.108$. Despite our somewhat crude assumptions, we find a very nice match for low multipole moments $l$ for both redshifts, only slightly overestimating the amplitude. For larger $l$ we find good agreement for $z=0.508$, slightly underestimating the power for the very largest $l$. For redshift $z=0.108$, the linear prediction overestimates the amplitude compared to the simulation result at large $l$. There are two possible explanations: i) Our window function was too imprecise, or ii) since these simulations have very little non-linear structure, with the initial spectrum even being cut off for very small scales to avoid numerical noise, it is likely that the simulation just underestimates the power here. This also explains why the non-linear signal we saw in our simulations was not picked up in the simulations of \cite{Koksbang2024}. Overall, the agreement between the simulation results of \cite{Koksbang2024} and our linear predictions for the power spectrum seems reasonable and gives us further confidence that our linear result is correct. 
\\ \\ 
The authors of \cite{Bessa2024} also calculated the redshift drift fluctuation power spectrum in a $\Lambda$CDM universe using N-body simulations. Their simulation was run using the Newtonian N-body code GADGET-4 \cite{Springel2021_GADGET4} and had a resolution and cosmological parameters similar to ours. The authors calculated the power spectrum from the peculiar velocities and acceleration on the constant time hypersurfaces of the simulation, i.e.\ using the background redshift and thus not raytracing to the individual simulation particles. Their power spectra disagree with ours in shape and amplitude\footnote{When comparing amplitudes, one has to account for the different normalizations employed by \cite{Bessa2024} and us. See the discussion in Appx.~\ref{appendix:Bessa2023}. This difference was taken into account in the comparison.}. The amplitudes in \cite{Bessa2024} also disagree with the perturbation theory results by \cite{Bessa2023}, but unfortunately the agreement is too large to be fixed by the disagreement we found with \cite{Bessa2023} in this paper, see Appx.~\ref{appendix:Bessa2023}. Since our simulation results agree very well with the predictions from perturbation theory and we have demonstrated that calculating the redshift drift from the peculiar velocity and acceleration is an excellent approximation, it seems most likely that the approximation of using the background redshift and not raytracing causes the discrepancies between the power spectra. 

\section{Implication for Observations}
\label{sec:obs_comparison}
Observations of the redshift drift with the SKA will be obtained by observing $\sim 10^7$ galaxies per redshift bin and stacking their signal, allowing a detection in the redshift range $0.2\lesssim z\lesssim 1$ \cite{Kloeckner2015}. More galaxies will be observed at lower redshifts, allowing a stronger detection at the lower end of this range. Due to the necessity of stacking many galaxies, such an experiment will constrain the mean/background signal of the redshift drift, averaging out fluctuations. Our results confirm that there is no significant bias to the mean from peculiar velocities or relativistic effects up to redshifts $z\approx 0.38$. When comparing the mean redshift drift of our simulations with the background value, we find only a very  minor deviation between the two across all ten considered observers. This strongly indicates that there is no significant observer-dependent bias from the large scale cosmic neighbourhood of the observer.
\\ \\
Observations of the redshift drift with the ELT will be obtained at much higher redshifts ($2\lesssim z\lesssim 5$) \cite{Liske2008} by studying quasar absorption lines. While our current simulations do not extend to such high redshifts, we have for the first time presented a numerical framework that can in principle accurately calculate the redshift drift signal including relativistic corrections (such as gravitational time dilation and metric perturbations) and light cone effects. This setup can be used to study the high redshift signal and check for potential biases e.g. due to ISW-like terms. Earlier estimates indicated that peculiar velocities will not bias the observed signal \cite{Liske2008, Cooke2020}. Explicitly demonstrating this with relativistic, non-linear simulations is an important step for ensuring reliable observational strategies and for building confidence in future detections. Due to the high computational cost, we leave a detailed high-redshift study to future work. Nonetheless, we have here already shown that at least at low redshifts, accounting for fluctuations from peculiar velocities only, and neglecting any relativistic corrections, allows us to reconstruct the redshift drift signal to better than one percent, independent of redshift. We expect that this would also be the case at high redshifts, except for potential integrated effects that grow with redshift, such as the ISW effect.
\\ \\
We also note that the strong non-linear signals we have described, while not observable with the current telescopes, could provide valuable insight into the gravitational potentials inside clusters and thereby provide strong constraints on modified theories of gravity. The prospect of detecting such a signal could provide motivation for future telescope missions, such as suggested in e.g. \cite{Eikenberry2019}.
\\ \\ 
Lastly we note that although the peculiar motion of sources has only small impact on expected redshift drift measurements with SKA and ELT, the effects may be more significant for alternative methods. This may for instance be the case for the proposed redshift difference which utilizes strong gravitational lensing to detect changes in the redshift due to different travel times \cite{Wang2022,Wang2023}.

\section{Conclusions}
\label{sec:conclusions}
The cosmological redshift drift promises to be the first observable that directly constrains the evolution of the cosmic expansion rate. To prepare these measurements and study the effect the structures in the universe will have on them, we have studied the redshift drift using general relativistic N-body simulations in a $\Lambda$CDM universe. We have calculated the redshift drift directly from the past light cone, without approximations, for 10 random observers.

Except at very low redshifts, $z\lesssim 0.05$, where we see a large scatter due to the local environment, the mean redshift drift follows the expected FLRW background signal with a relative deviation of approximately $0.05\%$ for all observers. For the largest redshifts in our simulation $0.28 \lesssim z \lesssim 0.38$, all observers see a smaller redshift drift than the background result. Since the difference is small, this is most likely not relevant for observations. 

Although the mean redshift drift closely follows the background result, fluctuations in the redshift drift can be of the same order of magnitude as the background signal for individual sources at redshifts $z\lesssim 0.15$. Such extreme outliers are due to sources with a high peculiar acceleration, sitting in high over-densities. These outliers may offer an opportunity to measure the gravitational potential of nearby galaxy clusters in the future. Whether or not this is actually feasible requires further investigation.

By comparing our simulation results with analytical expressions based on perturbation theory, we show that the fluctuations in the redshift drift can be described to better than $1\%$ accuracy for almost all particles in our simulation, accounting only for the peculiar acceleration and velocity of sources at the linear level, without the need to include metric perturbations. 

Lastly, we have calculated angular power spectra for the redshift drift fluctuations and compared them to predictions from linear perturbation theory. On linear scales, we find excellent agreement between the linear predictions and our simulation result in both shape and amplitude. This is contrary to the preprint \cite{Bessa2024} where significant differences between simulation and linear power spectra were found. Additionally, we found a surprisingly large non-linear signal arising from those sources with high peculiar acceleration and velocity, which has not been seen before, although several earlier studies have considered redshift drift fluctuation power spectra. 

Since different studies have used different definitions of fluctuations and/or used different background cosmologies, it has so far been difficult to compare existing results in the literature regarding redshift drift fluctuation power spectra. We remedy this by i) showing that our analytical predictions for the linear power spectra agree with earlier studies based on ET simulations with an EdS background, and ii) mapping between our linear results and those obtained earlier using a different definition of the redshift drift fluctuations, showing that our results disagree slightly with the earlier results in the numerical implementation, but not analytically.

\acknowledgments
AO would like to thank Queen Mary University of London for hospitality during part of this project and Asta Heinesen for valuable discussions. We thank Hayley J.\ Macpherson and Asta Heinesen for permission to reuse data from \cite{Koksbang2024}. We also thank the authors of \cite{Bessa2023} for sharing their modified version of CLASS and Pedro Bessa for discussing the results. 
SMK and AO are funded by VILLUM FONDEN, grant VIL53032. The simulations used in this paper were run using the UCloud interactive HPC system managed by the eScience Center at the University of Southern Denmark.
\newline\newline
{\bf Author contribution statement} 
AO performed all derivations and the numerical work with guidance from SMK and CC. JA provided guidance with respect to the numerical work. All authors have contributed to the interpretation of the results and the overall development of the project. AO led the writing of the paper, but all authors have contributed to the final manuscript.

\bibliography{main}

\begin{appendix}
\section{Comparison to Bessa et al 2023}
\label{appendix:Bessa2023}
\begin{figure}
    \centering
    \includegraphics[width=0.85\linewidth]{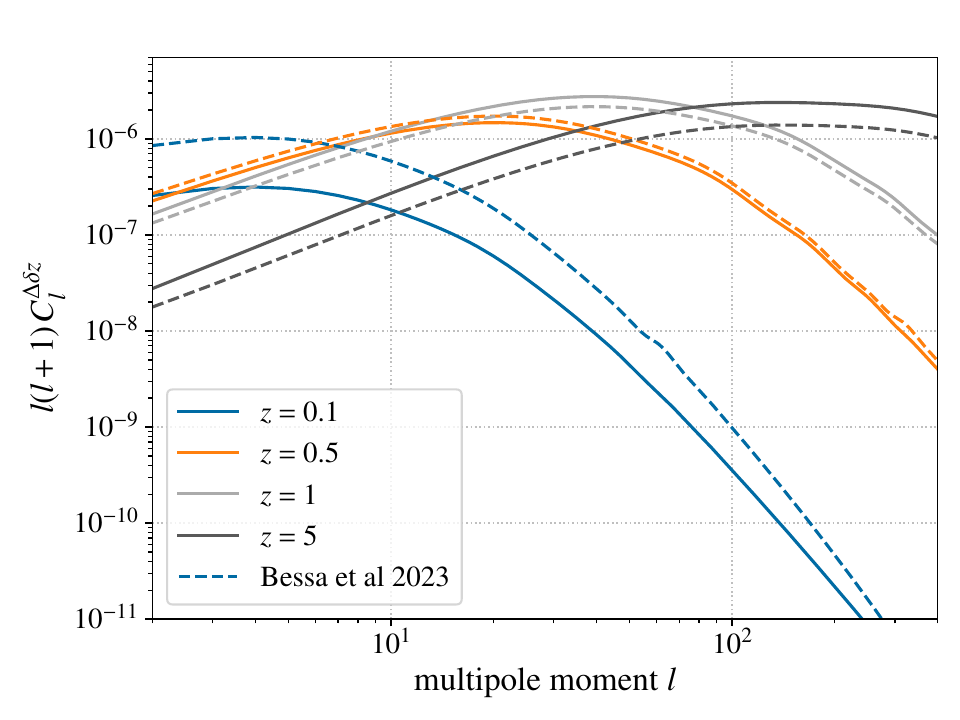}
    \caption{Comparison between predictions for the angular power spectrum of redshift drift fluctuations from perturbation theory as calculated in this paper and the results by \cite{Bessa2023}.}
    \label{fig:ps_comparison_bessa2023}
\end{figure}

In the main text we mention that the angular power spectrum derived in Sect.~\ref{sec:theory_ps} has earlier been derived in \cite{Bessa2023} in a different manner. The authors of \cite{Bessa2023} derived the spectrum in a fully gauge-independent manner and included also metric fluctuations, which we neglected here. According to our results in Sect.~\ref{sec:results_theor_pred} and also the conclusions in \cite{Bessa2023}, neglecting the metric perturbations is very well justified and should not influence the results. Adopting the convention used in \cite{Bessa2023}, the redshift drift fluctuations are represented as
\begin{equation}
	\Delta\delta z = \frac{\delta z - \delta \bar z}{\delta \bar z} =-\frac{\d}{\d t}\left( \frac{\bold v_\s}{H_\s}\right)\cdot \bold e\;,
    \label{eq:rsd_fluctuations_bessa23}
\end{equation}
i.e. neglecting the observer dependent term in \eqref{eq:rsd_fluctuations}. Neglecting this term only results in a different amplitude of the final spectra. Comparing \eqref{eq:rsd_fluctuations} and \eqref{eq:rsd_fluctuations_bessa23} one can read off that the final $C_l$'s differ by a factor 
\begin{equation}
    A=\left(\frac{H_\s}{(1+z)H_\o-H_s}\right)^2\;.
\end{equation}
Using the convention by \cite{Bessa2023}, we calculate the redshift drift fluctuation power spectrum once with our result \eqref{eq:rsd_Cl} and once with the CLASS modification implemented by \cite{Bessa2023}. We calculate the spectrum for redshifts $z=0.1,0.5,1,5$ with a Gaussian window function of width $\Delta z=0.01$, the same as in figure 3 of \cite{Bessa2023}. The result can be seen in Fig.~\ref{fig:ps_comparison_bessa2023}. One can see that there is a clear difference in amplitude between the two results, especially for low redshifts. Except the neglected potential perturbations, which \cite{Bessa2023} showed do not change their power spectrum, our analytical expression for the power spectrum are equivalent. The difference seen in Fig.~\ref{fig:ps_comparison_bessa2023} is therefore quite surprising. Since our results agree well with our simulation results, we are inclined to believe our implementation.

\section{Validation of the Power Spectra}
\label{appendix:ps_validation}
\begin{figure}
	\includegraphics[width=0.85\linewidth]{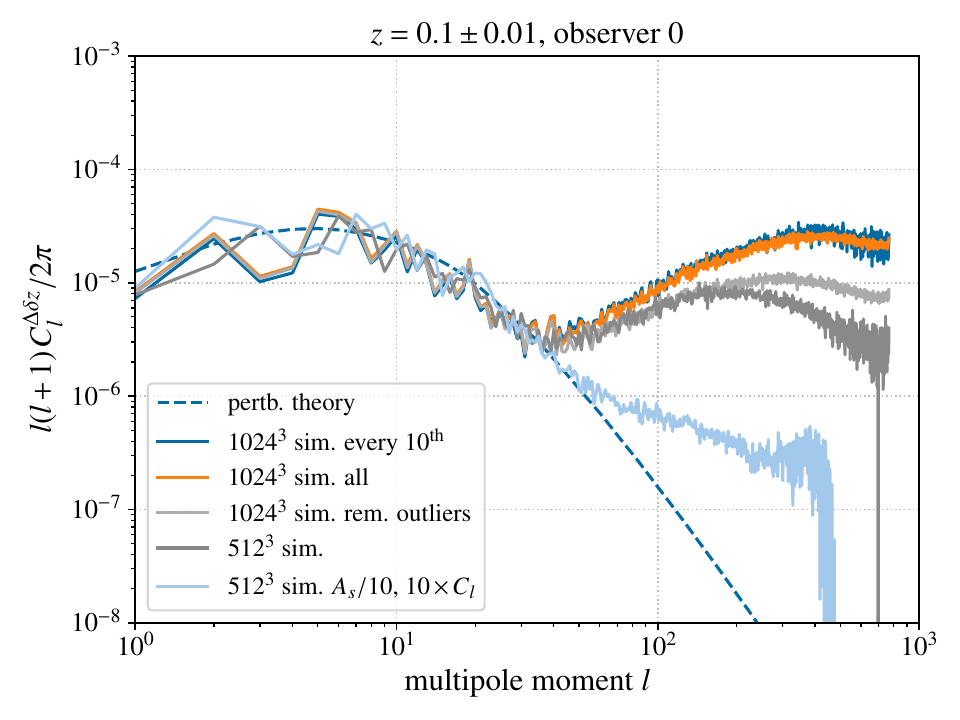}
	\caption{Power spectrum of the redshift drift fluctuations, for our main simulation  using all particles in the light cone (orange), only every tenth particle in the light cone (dark blue), having removed the $5\%$ furthest outliers from the mean redshift drift (light grey), a lower resolution simulation (dark grey), and the lower resolution with a lower initial amplitude of the primordial spectrum $A_s/10$. In the last case we plot $10\times C_l$ for easy comparison.}
    \label{fig:spectra_validation}
\end{figure}

In order to validate the power spectra we have shown in Sect.~\ref{sec:power_spectra} and make sure that the signal we are seeing at high $l$ is not numerical noise, we run a number of tests. We start by calculating the power spectrum with all of the particles in the light cone, rather than only every tenth, for one of our observers. The result can be seen in Fig.~\ref{fig:spectra_validation}. Comparing the dark blue and orange lines, we can see that while we get slightly better resolution for large $l$, including all the particles leaves the result overall unchanged. 

As a next test we run a simulation with half the resolution, i.e.\ $512^3$ particles and grid cells. In order to keep the same time-resolution we change the Courant factor to $c=1.5$. Since there are now fever particles in the light cone we use all of them to calculate the power spectrum. From a lower resolution simulation we would expect more noise, but less non-linear structure. The new power spectrum is shown as the dark grey line in Fig.~\ref{fig:spectra_validation}. We can see that the amplitude is now lower for large $l$, indicating that we are indeed picking up a signal from the non-linear structure and not numerical noise. 

We now also run the low resolution simulation with a lower amplitude of the primordial power spectrum used to set the initial conditions. Specifically, we now use $A_s/10$. Starting with a lower amplitude, i.e.\ less power, means that structures will take much longer to build up. In this case we now plot $10\times C_l$ such that the power spectra have the same amplitude for easy comparison. The resulting power spectrum is the light blue line in Fig.~\ref{fig:spectra_validation}. We can see that the result now follows the prediction from linear perturbation theory to larger $l$ and there is overall very little non-linear signal. 

As a last test we return to the higher resolution simulation and remove the $5\%$ of particles from the light cone with the largest distance to the mean, i.e.\ those with large peculiar accelerations in very non-linear environments. Calculating the power spectrum only with the remaining particles results in the light grey curve in Fig.~\ref{fig:spectra_validation}. We see that the non-linear signal is now reduced compared to the orange curve, again confirming that the signal is caused by the highly non-linear particles/environments in our simulation. We have also tried removing the $10\%$ furthest outliers. This simply reduced the amplitude on non-linear scales further, which we do not show again here. 

Using a larger HEALPix map resolution also resulted in no changes to the spectrum other than allowing us to go to slightly larger $l$. We only did this for one observer since it requires raytracing to all particles in the light cone to get sufficient resolution, which is numerically expensive. As there was not much new information we omit the plot here. 

\end{appendix}

\end{document}